\documentclass[prd,aps,showpacs,floats,letterpaper,floatfix,nofootinbib]{revtex4}

\usepackage{amsmath,amsbsy}
\usepackage{graphicx}
\usepackage{cancel}

\begin{document}

\title{Use and Abuse of the Fisher Information Matrix in the Assessment of Gravitational-Wave Parameter-Estimation Prospects}

\author{Michele Vallisneri}

\affiliation{Jet Propulsion Laboratory, California Institute of Technology, Pasadena, CA 91109}

\begin{abstract}
The Fisher-matrix formalism is used routinely in the literature on gravitational-wave detection to characterize the parameter-estimation performance of gravitational-wave measurements, given parametrized models of the waveforms, and assuming detector noise of known colored Gaussian distribution.
Unfortunately, the Fisher matrix can be a poor predictor of the amount of information obtained from typical observations, especially for waveforms with several parameters and relatively low expected signal-to-noise ratios (SNR), or for waveforms depending weakly on one or more parameters, when their priors are not taken into proper consideration. In this paper I discuss these pitfalls; show how they occur, even for relatively strong signals, with a commonly used template family for binary-inspiral waveforms; and describe practical recipes to recognize them and cope with them.

Specifically, I answer the following questions: (i) What is the significance of (quasi-)singular Fisher matrices, and how must we deal with them? (ii) When is it necessary to take into account prior probability distributions for the source parameters? (iii) When is the signal-to-noise ratio high enough to believe the Fisher-matrix result? In addition, I provide general expressions for the higher-order, beyond--Fisher-matrix terms in the 1/SNR expansions for the expected parameter accuracies.
\end{abstract}

\date{Jun 20 2007}
\pacs{04.80.Nn, 95.55.Ym, 02.50.Tt}

\maketitle

\section{Introduction}

Over the last two decades, the prevailing attitude in the gravitational-wave (GW) source-modeling community has been one of \emph{pre-data positioning}: in the absence of confirmed detections, the emphasis has been on exploring which astrophysical systems, and which of their properties, would become accessible to GW observations with the sensitivities afforded by planned (or desired) future experiments, with the purpose of committing theoretical effort to the most promising sources, and of directing public advocacy to the most promising detectors. In this positioning and in this exploration, the \emph{expected accuracy} of GW source parameters, as determined from the signals yet to be observed, is often employed as a proxy for the amount of physical information that could be gained from detection campaigns.  
However, predicting the parameter-estimation performance of future observations is a complex matter, even with the benefit of accurate theoretical descriptions of the expected waveforms and of faithful characterizations of the noise and response of detectors; in practice, the typical source modeler has had much less to go with. The main problem is that there are few analytical tools that can be applied generally to the problem, before resorting to relatively cumbersome numerical simulations that involve multiple explicit realizations of signal-plus-noise datasets.

In the source-modeling community, the analytical tool of choice has been the Fisher information matrix $F_{ij}[h] = (h_i,h_j)$: here $h_i(t)$ is the partial derivative of the gravitational waveform $h(t)$ of interest with respect to the $i$-th source parameter $\theta_i$, and ``$(\cdot,\cdot)$'' is a signal product weighted by the expected power spectral density of detector noise, as described in Sec.\ \ref{sec:gaussianlikelihood}. Now, it is usually claimed that the inverse Fisher matrix $F^{-1}_{ij}[h_0]$ represents the covariance matrix of parameter errors in the parameter-estimation problem for the true signal $h_0(t)$. This statement can be interpreted in three slightly different ways (all correct), which we examine in detail in Sec.\ \ref{sec:threeways}, and preview here:
\begin{enumerate}
\item The inverse Fisher matrix $F^{-1}_{ij}[h_0]$ is a lower bound (generally known as the \emph{Cram\'er--Rao bound}) for the error covariance of any \emph{unbiased estimator} of the true source parameters. Thus, it is a \emph{frequentist} error (see Sec.\ \ref{sec:freqnbayes}): for any \emph{experiment} characterized by the true signal $h_0(t)$ and a certain realization $n(t)$ of detector noise, the parameter estimator $\hat{\theta}$ is a vector function of the total detector output $s = n + h_0$, and $F^{-1}_{ij}[h_0]$ is a lower bound on the covariance (i.e., the fluctuations) of $\hat{\theta}$ in an imaginary infinite sequence of experiments with different realizations of noise. The Cram\'er--Rao bound is discussed in Sec.\ \ref{sec:cramerrao}.
\item The inverse Fisher matrix $F^{-1}_{ij}[h_0]$ is the frequentist error covariance for the \emph{maximum-likelihood} (ML) parameter estimator $\hat{\theta}^\mathrm{ML}$, assuming Gaussian noise, in the limit of strong signals (i.e., high signal-to-noise ratio SNR) or, equivalently, in the limit in which the waveforms can be considered as linear functions of source parameters. We shall refer to this limit as the linearized-signal approximation, or LSA. This well-known result is rederived in Sec.\ \ref{sec:frequentisthighsn}.
\item The inverse Fisher matrix $F^{-1}_{ij}[h_0]$ represents the covariance (i.e., the multidimensional spread around the mode) of the \emph{posterior probability distribution} $p(\theta_0|s)$ for the true source parameters $\theta_0$, as inferred (in \emph{Bayesian} fashion) from a \emph{single} experiment with true signal $h_0$, assuming Gaussian noise, in the high-SNR limit (or in the LSA), and in the case where any \emph{prior probabilities} for the parameters are constant over the parameter range of interest. Properly speaking, the inverse Fisher matrix is a measure of uncertainty rather than error, since in any experiment the mode will be displaced from the true parameters by an unknown amount due to noise.\footnote{In the high-SNR/LSA limit with negligible priors, the posterior probability mode, seen as a frequentist statistic, coincides with the ML estimator; thus its fluctuations are again described by the inverse Fisher matrix.} See Sec.\ \ref{sec:bayeshighsn} for a rederivation of this result.
\end{enumerate}
As pointed out by Jaynes \cite{jaynes2003}, while the numerical identity of these three different error-like quantities has given rise to much confusion, it arises almost trivially from the fact that in a neighborhood of its maximum, the signal likelihood $p(s|\theta_0)$ is approximated by a normal probability distribution with covariance $F^{-1}_{ij}$. In this paper, I argue that the Cram\'er--Rao bound is seldom useful in the work of GW analysts (Sec.\ \ref{sec:cramerrao}), and while the high-SNR/LSA frequentist and Bayesian results are legitimate, they raise the question of whether the signals of interest are strong (or linear) enough to warrant the limit, and of what happens if they are not. In addition, if we possess significant information about the prior distributions (or even the allowed ranges) of source parameters, it is really only in the Bayesian framework that we can fold this information reliably into the Fisher result (Sec.\ \ref{sec:frequentisthighsn}).

Thus, I recommend the Bayesian viewpoint as the most fruitful way of thinking about the Fisher-matrix result (although I will also derive parallel results from the frequentist viewpoint). Of course, the study of Bayesian inference for GW parameter-estimation problems need not stop at the leading-order (Fisher-matrix) expression for the posterior likelihood: Markov Chain Monte Carlo (MCMC) algorithms \cite{mcmc} can provide very reliable results, immune from any considerations about signal strength, but they require a significant investment of time to implement them, and of computational resources to run them, since they necessarily involve explicit realizations of the noise. More rigorous Bayesian bounds (such as the Weiss--Weinstein and Ziv--Zakai bounds examined by Nicholson and Vecchio \cite{nv1998}) can also be derived, but they require a careful appraisal of the nonlocal structure of the likelihood function.

By contrast, the Fisher-matrix formalism is singularly economical, and it seems clear that it will always be the first recourse of the GW data analyst. To use it reliably, however, we must understand the limits of its applicability. The purpose of this paper is to explore these limits. I do so by providing practical solutions to three issues that were already raised in the seminal treatments of GW detection by Finn \cite{finn1992} and by Cutler and Flanagan \cite{cf1994}, but that seem to have been almost ignored after that:
\begin{enumerate}
\item What is the significance of the singular or ill-conditioned Fisher matrices that often appear in estimation problems with several source parameters, and how do we deal with them? Can we still believe the Fisher result in those cases? (See Sec.\ \ref{sec:disappearing}.)
\item When is it necessary to take into account the prior probability distributions for the parameters, even if specified trivially by their allowed ranges? (See Sec.\ \ref{sec:priors}.)
\item When is the high-SNR/LSA approximation warranted? (As anticipated above, the high-SNR limit is \emph{equivalent} to the LSA, as we shall show in Secs.\ \ref{sec:frequentisthighsn} and \ref{sec:bayeshighsn}.) That is, how strong a signal will we need to measure if we are to believe the Fisher-matrix result for its uncertainty? (See Sec.\ \ref{sec:howhigh}.)
\end{enumerate}
Last, I discuss the extension of the LSA beyond the leading order, in both the frequentist and Bayesian parameter-estimation frameworks (Sec.\ \ref{sec:higherorder}), in a form that the adventurous GW analyst can use to test the reliability of the Fisher result (but higher-order derivatives and many-indexed expressions start to mount rapidly, even at the next-to-leading order). 
By contrast, I do not address the reduction in parameter-estimation accuracy due to the presence of secondary maxima in the likelihood function, as noticed \cite{bsd1996} and carefully modeled \cite{bd1998} by Balasubramanian and colleagues in their extensive Monte Carlo simulations of ML estimation for inspiraling binaries using Newtonian and first post-Newtonian waveforms.

My treatment follows Refs.\ \cite{finn1992,cf1994}, as well as the classic texts on the statistical analysis of noisy data (e.g., Refs.\ \cite{wainstein62,oppenheim83,kay1993}). I am indebted to Jaynes and Bretthorst \cite{jaynes2003,brett1988} for their enlightening, if occasionally blunt, perspective on frequentist and Bayesian parameter estimation. The reader already familiar with the standing of the Fisher-matrix formalism in the frequentist and Bayesian frameworks can skip Secs.\ \ref{sec:freqnbayes} (a refresher on the difference between the frequentist and Bayesian viewpoints) and \ref{sec:cramerrao}--\ref{sec:bayeshighsn} (a pedagogical derivation of the three approaches to the inverse--Fisher-matrix result that were introduced at the beginning of this section), and move directly to discussion of the three \emph{issues} in Secs.\ \ref{sec:disappearing}--\ref{sec:howhigh}, and to the higher-order formalism in Sec.\ \ref{sec:higherorder}, referring back to Sec.\ \ref{sec:threeways} as needed to establish notation.
Whenever my discussion requires a practical example, I consider signals from inspiraling binaries of two black holes, both of mass $10 M_\odot$, as described by the restricted post-Newtonian approximation for adiabatic, circular inspirals (see Sec.\ \ref{sec:standardsignal}); in my examples, I assume detection and parameter estimation are performed on Initial-LIGO \cite{ligo} data, and I adopt the LIGO noise curve of Table IV in Ref.\ \cite{DIS3}. Throughout, I use geometric units; I assume the Einstein summation convention for repeated indices; and I do not distinguish between covariant and contravariant indices, except in Sec.\ \ref{sec:higherorder}.

\section{Three roads to the Fisher matrix}
\label{sec:threeways}

In this section I discuss the ``three roads'' to the inverse Fisher matrix as a measure of uncertainty for GW observations: the Cram\'er--Rao bound (Sec.\ \ref{sec:cramerrao}), the high-SNR/LSA limit for the frequentist covariance of the ML estimator (Sec.\ \ref{sec:frequentisthighsn}), and the high-SNR/LSA limit for the single-experiment covariance of the Bayesian posterior distribution (Sec.\ \ref{sec:bayeshighsn}). Sections \ref{sec:freqnbayes} and \ref{sec:gaussianlikelihood} are refreshers about frequentist and Bayesian parameter estimation, and about the analytical expression for the likelihood of GW signals in Gaussian noise.

\subsection{A refresher on the frequentist and Bayesian frameworks}
\label{sec:freqnbayes}

\noindent The \emph{frequentist} (or orthodox) approach to parameter estimation for GW signals can be summed up as follows:
\begin{enumerate}
\item We are given the detector data $s$ and we take it to consist of the true signal $h_0 = h(\theta_0)$ (where $\theta_0$ is the vector of the \emph{true system parameters}) plus additive noise $n$.
\item We select a \emph{point estimator} $\hat{\theta}(s)$: that is, a vector function of detector data that (it is hoped) approximates the true values of source parameters, except for the statistical error due to the presence of noise. One important example of point estimator is the ML estimator $\hat{\theta}^\mathrm{ML}$, which maximizes the \emph{likelihood} $p(s|\theta)$ of observing the measured data $s$ given a value $\theta$ of the true parameters. For additive noise, this likelihood coincides with the probability of a noise realization $n = s - h(\theta)$, and for Gaussian noise it is given below in Sec.\ \ref{sec:gaussianlikelihood}.
\item We characterize statistical error as the fluctuations of $\hat{\theta}(s)$, computed over a very long series of independent \emph{experiments} where the source parameters are kept fixed, while detector noise $n$ is sampled from its assumed probability distribution (often called the \emph{sampling} distribution).
\end{enumerate}
The estimator $\hat{\theta}$ is usually chosen according to one or more criteria of optimality: for instance, \emph{unbiasedness} requires that $\langle \hat{\theta}(s) \rangle_n$ (the average of the estimator over the noise probability distribution) be equal to $\theta_0$.

A rather different approach is that of \emph{Bayesian} inference:
\begin{enumerate}
\item We do not assume a true value of the system parameters, but we posit their \emph{prior probability distribution} $p(\theta)$.
\item Given the data $s$, we do not compute estimators, but rather the full \emph{posterior probability distribution} $p(\theta|s)$, using \emph{Bayes' theorem} $p(\theta|s) = p(s|\theta) \times p(\theta) / p(s)$, where $p(s) = \int p(s|\theta) \, p(\theta) \, d\theta$.
\item We characterize statistical error \emph{in a single experiment} by the spread of the posterior distribution $p(\theta|s)$.
\end{enumerate}
The differences between the frequentist and Bayesian approaches are not only mathematical, but also epistemic: as their name indicates, ``frequentists'' view probabilities essentially as the relative frequencies of outcomes in repeated experiments, while ``Bayesians'' view them as subjective\footnote{Only in the sense that subjects with different prior assumptions could come to different conclusions after seeing the same data; indeed, Bayesian statistics describes how prior assumptions become \emph{deterministically} modified by the observation of data.} indices of certainty for alternative propositions. For an introduction to the contrasting views, I refer the reader to the excellent treatise (very partial to the Bayesian worldview) by Jaynes \cite{jaynes2003}, and to Ref.\ \cite{cf1994} for a more GW-detection--oriented discussion.

Once actual detections are made, the Bayesian approach of computing posterior probability distributions for the signal parameters \emph{given the observed data} seems more powerful than the frequentist usage of somewhat arbitrary point estimators; the latter will always result in throwing away useful information, unless the chosen estimators are \emph{sufficient statistics} (i.e., unless the likelihood depends on the data only through the estimators). As for statistical error, it seems preferable to characterize it from the data \emph{we have} (actually, from the posterior distributions that we infer from that data), rather than from the data \emph{we could have obtained} (i.e., from the sampling distribution of estimators in a hypothetical \emph{ensemble} of experiments).

As Cutler and Flanagan \cite{cf1994} point out, however, it is in the current \emph{pre-data} regime that we seek to compute expected parameter accuracies; in the absence of actual confirmed-detection datasets, it seems acceptable to consider \emph{ensembles} of possible parameter-estimation experiments, and to use frequentist statistical error as an inverse measure of potential physical insight. The best solution, bridging the two approaches, would undoubtedly be to examine the frequentist distribution of some definite measure of Bayesian statistical error; unfortunately, such a hybrid study is generally unfeasible, given the considerable computational requirements of even single-dataset Bayesian analyses.

\subsection{Likelihood for GW signals in Gaussian noise}
\label{sec:gaussianlikelihood}

Under the assumption of stationary and Gaussian detector noise, the likelihood $\log p(s|\theta)$ can be obtained very simply from a noise-weighted inner product of the detector output and of the signal $h(\theta)$ (see for instance Eq.\ (2.3) in Ref.\ \cite{cf1994}):
\begin{equation}
p(s|\theta) \propto e^{-(s-h(\theta),s-h(\theta))/2};
\label{eq:pstheta}
\end{equation}
the weighting is performed with respect to the expected power spectral density of detector noise by defining the 
noise-weighted inner product of two real-valued signals as
\begin{equation}
\label{eq:noiseproduct}
(h,g) = 4 \, \mathrm{Re} \int_0^{+\infty} \frac{\tilde{h}(f)^* \tilde{g}(f)}{S_n(f)} \, df,
\end{equation}
where $\tilde{h}(f)$ and $\tilde{g}(f)$ are the Fourier transforms of $h(t)$ and $g(t)$, ``$*$'' denotes complex conjugation, and $S_n(f)$ is the one-sided power spectral density of the noise. From the definition of $S_n(f)$ as $\langle \tilde{n}^*(f) \tilde{n}(f') \rangle_n = \frac{1}{2} S_n(|f|) \, \delta(f - f')$, we get the useful property
\begin{equation}
\label{eq:noiseprod}
\bigl\langle
(h,n) (n,g)
\bigr\rangle_n = (h,g),
\end{equation}
where again ``$\langle \cdot \rangle_n$'' denotes averaging over the probability distribution of the noise.

\subsection{First road: Derivation and critique of the Cram\'er--Rao bound}
\label{sec:cramerrao}

The derivation in this section is inspired by the treatment of Ref.\ \cite[p.\ 518]{jaynes2003}, and it is given for simplicity in the case of one source parameter. We wish to pose a bound on the frequentist estimator variance
\begin{equation}
\mathrm{var} \, \hat{\theta} =
\Bigl\langle \Bigl( \hat{\theta}(s) - \bigl\langle \hat{\theta}(s) \bigr\rangle \Bigr)^2 \Bigr\rangle_n:
\end{equation}
to do this, we consider the ensemble product
\begin{equation}
\bigl\langle u(s), v(s) \bigr\rangle_n = \int u(s) \, v(s) \, p(s|\theta_0) \, ds,
\end{equation}
where $p(s|\theta_0)$ is the likelihood of observing the detector output $s$ given the true source parameter $\theta_0$, or equivalently the likelihood of observing the noise realization $n = s - h_0$. Setting $v(s) = \hat{\theta}(s) - \bigl\langle \hat{\theta}(s) \bigr\rangle_n$, we obtain a bound on $\langle v,v \rangle_n \equiv \mathrm{var}\, \hat{\theta}$ from the Schwarz inequality:
\begin{equation}
\mathrm{var} \, \hat{\theta} \equiv \langle v, v \rangle_n \geq \frac{\langle u, v \rangle_n^2}{\langle u, u \rangle_n}.
\label{eq:schwarz}
\end{equation}
This inequality is true for any function $u(s)$ of the data, and it becomes an equality when $u(s) \propto v(s)$.
Since we wish to derive a bound that applies generally to all estimators, we should not have (or try) to provide too much detail about $\hat{\theta}$ (and therefore $v(s)$). A simple assumption to make on $\hat{\theta}$ is that it is an unbiased estimator:
\begin{equation}
\bigl\langle \hat{\theta}(s) \bigr\rangle_n = \theta_0
\quad \Rightarrow \quad
\partial_{\theta_0} \bigl\langle \hat{\theta}(s) \bigr\rangle_n = 1.
\end{equation}
How does this help us? It turns out that we can write a function $d(s)$ whose ensemble product with any other function $w(s)$ yields the derivative $\partial_{\theta_0} \langle w(s) \rangle_n$; this function is just $d(s) = \partial_{\theta_0} \log p(s|\theta_0)$, because
\begin{equation}
\int w(s) [\partial_{\theta_0} \log p(s|\theta_0)] \, p(s|\theta_0) \, ds = 
\int w(s) \, \partial_{\theta_0} p(s|\theta_0) \, ds =
\partial_{\theta_0} \int w(s) p(s|\theta_0) \, ds =
\partial_{\theta_0} \langle w(s) \rangle_n,
\label{eq:dercramer}
\end{equation}
assuming of course\footnote{This assumption fails for some (mildly) pathological likelihood functions, which can provide counterexamples to the Cram\'er--Rao bound.} that we can exchange integration and differentiation with respect to $\theta_0$.
For any $s$, $d(s)$ encodes the \emph{local relative change} in the likelihood function as $\theta_0$ is changed. It follows that $\langle d(s), v(s) \rangle_n = \partial_{\theta_0} \bigl\langle \hat{\theta}(s) \bigr\rangle_n = 1$, so from Eq.\ \eqref{eq:schwarz} we get\footnote{To obtain Eq.\ \eqref{eq:cramersimple}, we need to notice also that for any $w(s)$, $\langle d(s), \langle w(s) \rangle_n \rangle_n = 0$, since $\langle w(s) \rangle_n$ does not depend on $s$ (but only on $\theta_0$), and the integral of Eq.\ \eqref{eq:dercramer} reduces to $\langle w(s) \rangle_n \int \partial_{\theta_0} p(s|\theta_0) ds = \langle w(s) \rangle_n \partial_{\theta_0} 1 = 0$.}
\begin{equation}
\mathrm{var} \, \hat{\theta} \geq \frac{1}{\bigl\langle d(s), d(s) \bigr\rangle}
\equiv \frac{1}{\bigl\langle \partial_{\theta_0} \log p(s|\theta_0) , \partial_{\theta_0} \log p(s|\theta_0) \bigr\rangle_{n}},
\label{eq:cramersimple}
\end{equation}
which is the unbiased-estimator version of the Cram\'er--Rao bound. If the estimator is biased, we can still use the Schwarz inequality by providing the derivative of the \emph{bias} $b(\theta_0)$ with respect to $\theta_0$:
\begin{equation}
\langle \hat{\theta}(s) \rangle_n = \theta_0 + b(\theta_0)
\quad \Rightarrow \quad
\partial_{\theta_0} \langle \hat{\theta}(s) \rangle_n = 1 + \partial_{\theta_0} b(\theta_0),
\end{equation}
and therefore
\begin{equation}
\label{eq:cramersingle}
\mathrm{var} \, \hat{\theta} \geq \frac{(1 + \partial_{\theta_0} b)^2}{\bigl\langle \partial_{\theta_0} \log p(s|\theta_0), \partial_{\theta_0} \log p(s|\theta_0) \bigr\rangle_{n}}.
\end{equation}
Generalizing to a multidimensional expression is straightforward, if verbose (see, e.g., Ref.\ \cite{kay1993}):
\begin{equation}
\mathrm{covar}_n(\hat{\theta}_i,\hat{\theta}_l) \geq
\bigl(\delta_{im} + \partial_m b_i(\theta_0)\bigr) F^{-1}_{mj} \bigl(\delta_{jl} + \partial_j b_l(\theta_0)\bigr),
\label{eq:cramerfull}
\end{equation}
where the \emph{Fisher information matrix} is defined by
\begin{equation}
F_{il} = \Bigl\langle \bigl(\partial_i \log p(s|\theta_0)\bigr), \bigl(\partial_l \log p(s|\theta_0)\bigr) \Bigr\rangle_{n} = 
- \Bigl\langle \partial_i \partial_l \log p(s|\theta_0) \Bigr\rangle_{n}\,.
\label{eq:fisherdef}
\end{equation}
The second equality is established by taking the gradient of $\int (\partial_i \log p(s|\theta_0)) \, p(s|\theta_0) \, ds$, and remembering that $\partial_i \int p(s|\theta_0) \, ds = \partial_i 1 = 0$. With the help of Eqs.\ \eqref{eq:pstheta} and \eqref{eq:noiseprod}, we can compute the Fisher matrix for GW signals in additive Gaussian noise, which is the familiar expression $F_{ij} = (\partial_i h,\partial_j h)$.

The full expression \eqref{eq:cramerfull} for the Cram\'er--Rao bound, which includes the effects of bias, has interesting consequences, for it implies that biased estimators can actually \emph{outperform}\footnote{This is true even if we evaluate the performance of estimators on the basis of their quadratic error
\begin{displaymath}
\langle (\hat{\theta}_i - \theta_{0i}) (\hat{\theta}_l - \theta_{0l}) \rangle_n \geq
b_i(\theta_0) b_l(\theta_0) + \bigl(\delta_{im} + \partial_m b_i(\theta_0)\bigr) F^{-1}_{mj} \bigl(\delta_{jl} + \partial_j b_l(\theta_0)\bigr)
\end{displaymath}
rather than on the basis of their variance.} unbiased estimators, since the $\partial_m b_i(\theta_0)$ can be negative. Unfortunately, we have no handle on these derivatives without explicitly choosing a particular estimator (which goes against the idea of having a generic \emph{bound}), so the Cram\'er--Rao bound can only give us a definite result for the subclass of unbiased estimators.

As pointed out by Cutler and Flanagan \cite[App.\ A 5]{cf1994}, it follows that the bound cannot be used to place absolute limits on the accuracy of estimators (i.e., lower bounds on frequentist error)---limits that would exclude or severely limit the possibility of inferring the physical properties of sources from their emitted GWs. Even if the lower bound for unbiased estimators is very discouraging, there is always a chance that a biased estimator could do much better, so we cannot use the bound to prove ``no go'' theorems.

Going back to Eq.\ \eqref{eq:schwarz}, we note that the bound is satisfied as an equality when 
\begin{equation}
\label{eq:exponentialequation}
u(s) \propto v(s)
\quad \Rightarrow \quad
d(s) \equiv \partial_{\theta_0} \log p(s|\theta_0) = q(\theta_0) [\hat{\theta}(s) - \langle \hat{\theta}(s) \rangle_n].
\end{equation}
By integrating, we obtain a relation between the likelihood and the estimator:
\begin{equation}
\label{eq:exponentialfamily}
p(s|\theta_0) = \frac{m(s)}{Z(\theta_0)} e^{-l(\theta_0) \hat{\theta}(s)};
\end{equation}
the estimation problems (i.e., the pairings of given likelihoods and chosen estimators) for which this relation holds true are said to belong to the \emph{exponential family}, and these problems are the only ones for which the Cram\'er--Rao bound is satisfied exactly as an equality. Equation \eqref{eq:exponentialfamily} generalizes trivially to multidimensional problems by replacing the exponential with $\exp \, \{-l_k(\theta_0) \hat{\theta}_k(s)\}$. Unfortunately, for a given $p(s|\theta_0)$ there is no guarantee that any unbiased estimator exists that satisfies Eq.\ \eqref{eq:exponentialfamily} and that therefore can actually \emph{achieve} the bound; all we can say in general about the performance of \emph{unbiased} estimators is that they will underperform the Cram\'er--Rao bias, but we do not know how badly. As discussed above, the bound tells us nothing in general about \emph{biased} estimators.

It follows that the bound cannot be used to establish guaranteed levels of accuracy (i.e., upper bounds on frequentist error), which would prove the possibility of inferring the physical properties of sources from their GWs. We can only do so if we can identify a specific estimator that achieves the bound. In the next section we shall see that the ML estimator\footnote{Indeed, Eq.\ \eqref{eq:exponentialfamily} implies that if both an \emph{efficient} (i.e., bound-achieving) unbiased estimator and the ML estimator exist, they must coincide. To show this, we notice that if the ML estimator exists, the log-derivative $\partial_i \log p(s|\theta) = -\partial_i l_k(\theta) (\hat{\theta}_k - \theta_k)$ must be zero at $\theta = \hat{\theta}^\mathrm{ML}$, from which it follows that $\hat{\theta}_k = \hat{\theta}^\mathrm{ML}_k$.} does so in the high-SNR limit, where waveforms can be approximated accurately as \emph{linear} functions of their parameters within the region of parameter space where $p(s|\theta)$ is not negligible (so the high-SNR limit coincides with the limit in which the LSA is accurate).

We conclude that the Cram\'er--Rao bound is seldom useful to the GW analyst as a proper bound, whether to make positive or negative expected-accuracy statements; where it is useful, it reduces to the high-SNR/LSA result for the ML estimator.

\subsection{Second road: Derivation and critique of the frequentist high-SNR/LSA result}
\label{sec:frequentisthighsn}

We denote the true signal as $h_0$ (so $s = h_0 + n$), and expand the generic waveform $h(\theta)$ around $h_0$, normalizing signals by the \emph{optimal signal-to-noise ratio} of the true signal, $A = \sqrt{(h_0,h_0)}$ (also known in this context as \emph{signal strength}):
\begin{equation}
h(\theta) = h_0 + \theta_k h_k + \theta_j \theta_k h_{jk} / 2 + \cdots 
= A (\bar{h}_0 + \theta_k \bar{h}_k + \theta_j \theta_k \bar{h}_{jk} / 2 + \cdots);
\end{equation}
here we are translating source parameters as needed to have $h(0) = h_0$, defining $h_i = \partial_i h|_{\theta = 0}$, $h_{ij} = \partial_{ij} h|_{\theta = 0}$ (and so on), and $\bar{h}_0 = h_0 / A$, $\bar{h}_k = h_k / A$ (and so on).\footnote{The statistical uncertainty in the estimated signal strength can still be handled in this notation by taking one of the $\bar{h}_k$ to lie along $\bar{h}_0$; the corresponding $\theta_k$ represents a fractional correction to the true $A$.}
The likelihood is then given by\footnote{Formally, it is troubling to truncate the series expression for the exponent at any order beyond quadratic, since the integral of the truncated likelihood may become infinite; the important thing to keep in mind, however, is that the series need converge only within a limited parameter range determined self-consistently by the truncated-likelihood estimator, by compact parameter ranges, or (in the Bayesian case) by parameter priors. Similar considerations apply to the derivation of the higher-order corrections given in Sec.\ \ref{sec:higherorder}.}
\begin{equation}
\begin{aligned}
\label{eq:likexpanded}
p(s|\theta) \propto e^{-(s - h(\theta),s - h(\theta))/2} =
\exp \bigr\{ & -(n,n)/2 - A^2 \bigl[ \theta_j \theta_k (\bar{h}_j,\bar{h}_k) + \theta_j \theta_k \theta_l (\bar{h}_j,\bar{h}_{kl}) + \cdots \bigr] / 2  \\
 & + A \bigl[\theta_j (n,\bar{h}_j) + \theta_j \theta_k (n,\bar{h}_{jk}) / 2 + \theta_j \theta_k \theta_l  (n,\bar{h}_{jkl}) / 3! + \cdots \bigr] \bigr\}.
\end{aligned}
\end{equation}
The ML equations $\partial_j p(s|\theta^\mathrm{ML}) = 0$ are given by
\begin{equation}
\label{eq:mlexpanded}
0 = \frac{1}{A} \left[
(n,\bar{h}_j) + \hat{\theta}_k (n,\bar{h}_{jk}) + \hat{\theta}_k \hat{\theta}_l (n,\bar{h}_{jkl}) / 2 + \cdots \right] - \left[
\hat{\theta}_k (\bar{h}_j,\bar{h}_k)
+ \hat{\theta}_k \hat{\theta}_l \left(
(\bar{h}_j,\bar{h}_{kl})/2 + (\bar{h}_k,\bar{h}_{jl})
\right) + \cdots
\right],
\end{equation}
where we have divided everything by $A^2$, and we omit the ``${}^\mathrm{ML}$'' superscript for conciseness. A careful study of Eq.\ \eqref{eq:mlexpanded} shows that
it can be solved in perturbative fashion by writing $\hat{\theta}_j^\mathrm{ML}$
as a series in $1/A$,
\begin{equation}
\label{eq:mlestimatorexpand}
\hat{\theta}_j^\mathrm{ML} = 
\hat{\theta}_j^{(1)} / A + \hat{\theta}_j^{(2)} / A^2 + \hat{\theta}_j^{(3)} / A^3 + \cdots,
\end{equation}
and by collecting the terms of the same order in Eq.\ \eqref{eq:mlexpanded},
\begin{equation}
\label{eq:expandsystem}
\begin{aligned}
O(1/A):   \quad & (n,\bar{h}_j) - \hat{\theta}_k^{(1)} (\bar{h}_j,\bar{h}_k) = 0, \\
O(1/A^2): \quad & \hat{\theta}_k^{(1)} (n,\bar{h}_{jk}) 
- \hat{\theta}_k^{(1)} \hat{\theta}_l^{(1)} \left(
(\bar{h}_j,\bar{h}_{kl})/2 + (\bar{h}_k,\bar{h}_{jl})
\right)
- \hat{\theta}_k^{(2)} (\bar{h}_j,\bar{h}_k) = 0, \\
O(1/A^3): \quad & \ldots
\end{aligned}
\end{equation}
thus the ML solution $\hat{\theta}_j^\mathrm{ML}$ is given by 
\begin{equation}
\label{eq:solbyorders}
\begin{aligned}
\hat{\theta}_j^\mathrm{ML} = \, & \frac{1}{A} \, (\bar{h}_j,\bar{h}_k)^{-1} (\bar{h}_k,n) \, + \\
& \frac{1}{A^2} \,
(\bar{h}_j,\bar{h}_i)^{-1} \bigl\{
(n,\bar{h}_{ik}) (\bar{h}_k,\bar{h}_l)^{-1} (\bar{h}_l,n) 
- \left(
(\bar{h}_i,\bar{h}_{kl})/2 + (\bar{h}_k,\bar{h}_{il})
\right)
(\bar{h}_k,\bar{h}_m)^{-1} (\bar{h}_m,n)
(\bar{h}_l,\bar{h}_n)^{-1} (\bar{h}_n,n)
\bigr\} \, + \\
& \frac{1}{A^3} \, \bigl\{ \cdots \bigr\} + \cdots
\end{aligned}
\end{equation}

Thus we see that the limit of large $A$ (i.e., high SNR) coincides with the \emph{linearized-signal approximation} (LSA) where only the first derivatives of the signals are included. In the LSA, the likelihood is just
\begin{equation}
\begin{aligned}
\label{eq:standardlikelihood}
p(s|\theta) &\propto \exp \bigl\{ -(n,n)/2 - \theta_j \theta_k (h_j,h_k)/2 + \theta_j (h_j,n) \bigr\} \\
&= \exp \bigl\{ -(n,n)/2 - A^2 \theta_j \theta_k (\bar{h}_j,\bar{h}_k)/2 + A \theta_j \,(\bar{h}_j,n) \bigr\} \quad \text{(LSA)},
\end{aligned}
\end{equation}
and the ML estimator is given by
\begin{equation}
\label{eq:lsaest}
\hat{\theta}_j^\mathrm{ML} = (1/A) (\bar{h}_j,\bar{h}_k)^{-1} (\bar{h}_k,n) \quad \text{(LSA)}.
\end{equation}
Since $\langle(\bar{h}_k,n)\rangle_n = 0$, we see also that the ML estimator is unbiased.
The variance of $\hat{\theta}^\mathrm{ML}$ is then obtained by averaging  
$\hat{\theta}_j^\mathrm{ML} \hat{\theta}_k^\mathrm{ML}$ over noise realizations,
\begin{equation}
\label{eq:lsavar}
\begin{aligned}
\bigl< \hat{\theta}_j^\mathrm{ML} \hat{\theta}_k^\mathrm{ML} \bigr>_n & =
\frac{1}{A^2} \, (\bar{h}_j,\bar{h}_l)^{-1} \bigl< (\bar{h}_l,n) (n, \bar{h}_m,n) \bigr>_n
(\bar{h}_m,\bar{h}_k)^{-1} = \\
&= \frac{1}{A^2} \, (\bar{h}_j,\bar{h}_l)^{-1} (\bar{h}_l,\bar{h}_m)
(\bar{h}_m,\bar{h}_k)^{-1} = \frac{1}{A^2} \, (\bar{h}_j,\bar{h}_k)^{-1} \quad \text{(LSA)},
\end{aligned}
\end{equation}
and it coincides with the mean quadratic error in the frequentist sense. In Eq.\ \eqref{eq:lsavar},
the second equality follows from Eq.\ \eqref{eq:noiseprod}. The interpretation of the limit is that, for strong signals, the typical $\hat{\theta}^\mathrm{ML}_j - \theta_{0j}$ becomes small enough that the log-likelihood is accurately described by the product of detector data and a linearized signal.

Equation \eqref{eq:lsavar} is the standard Fisher-information--matrix result, and it implies that in the high-SNR/LSA limit the ML estimator achieves the Cram\'er--Rao bound. As we shall see in Sec.\ \ref{sec:higherorder}, the next-order correction to the variance scales as $1/A^4$, not $1/A^3$. This is because all $O(1/A^3)$ terms contain odd numbers of $n$, whose products vanish under the ensemble average. The fact itself that there is a next-order correction shows that for generic $A$ the ML estimator does not achieve the bound.

The fact that the Cram\'er--Rao bound is achieved in the high-SNR/LSA limit, but not beyond it, can also be seen in the light of Eq.\ \eqref{eq:exponentialfamily}, which encodes a standard form for the estimation problems in the exponential family. To express the LSA likelihood in this form, we can set $m(s) = e^{-(s - h_0,s - h_0)/2}$ and $Z(\theta) = e^{\theta_j \theta_k (h_j,h_k)/2}$; it remains to establish that
\begin{equation}
-l_j(\theta) \hat{\theta}^\mathrm{ML}_j(s) = \theta_j (h_j, s - h_0),
\end{equation}
which is satisfied by $l_j(\theta) = -(h_j,h_k) \theta_k$ [see Eq.\ \eqref{eq:lsaest}]. Now, if additional terms are added to Eq.\ \eqref{eq:standardlikelihood}, beginning with terms cubic in the $\theta_i$, $\hat{\theta}_j^\mathrm{ML}(s)$ comes to be a nonlinear function of the signal, such that no $-l_j(\theta)$ can multiply it in the right way to reconstruct the likelihood. It then follows that the estimation problem moves outside the exponential family, and the Cram\'er--Rao bound cannot be achieved.

It is possible (but perhaps not desirable, as we shall see shortly) to modify the ML estimator to take into account prior knowledge about the expected distribution of sources. The resulting maximum-posterior estimator $\hat{\theta}^\mathrm{MP}$ is defined as the mode of the posterior probability $p(\theta|s) = p(s|\theta) p(\theta) / p(s)$,
\begin{equation}
\hat{\theta}^\mathrm{MP} = \mathrm{maxloc}_\theta \, p(\theta|s)
= \mathrm{maxloc}_\theta \, p(s|\theta) p(\theta).
\end{equation}
This is a biased estimator: in the high-SNR/LSA limit, and with a Gaussian prior $p(\theta) \propto \exp \{-P_{ij} (\theta_i - \theta^P_i) (\theta_j - \theta^P_j) / 2\}$ centered at $\theta^P$ (the only prior that can be easily handled analytically), we find
\begin{equation}
\label{eq:biaswithprior}
b^\mathrm{MP}_i = \left\langle \theta_i^\mathrm{MP} \right\rangle_n = 
\left[(\bar{h}_i,\bar{h}_j) + P_{ij}/A^2\right]^{-1} (P_{jk}/A^2) \, \theta^P_k \quad \text{(LSA/Gaussian prior)};
\end{equation}
thus the $\hat{\theta}^\mathrm{MP}$ becomes unbiased for $A \rightarrow \infty$ (indeed, in that limit $\hat{\theta}^\mathrm{MP}$ tends to $\hat{\theta}^\mathrm{ML}$). For the frequentist variance we find
\begin{equation}
\label{eq:varwithprior}
\begin{aligned}
\bigl\langle \hat{\theta}_i^\mathrm{MP} \hat{\theta}_j^\mathrm{MP} \bigr\rangle_n -
\bigl\langle \hat{\theta}_i^\mathrm{MP} \bigr\rangle_n
\bigl\langle \hat{\theta}_j^\mathrm{MP} \bigr\rangle_n & =
\bigl\langle \hat{\theta}_i^\mathrm{MP} \hat{\theta}_j^\mathrm{MP} \bigr\rangle_n - b^\mathrm{MP}_i b^\mathrm{MP}_j \\
& = \frac{1}{A^2}
\left[(\bar{h}_i,\bar{h}_k) + P_{ik} / A^2 \right]^{-1} (\bar{h}_k,\bar{h}_l)
\left[(\bar{h}_l,\bar{h}_j) + P_{lj} / A^2 \right]^{-1} \quad \text{(LSA w/prior)},
\end{aligned}
\end{equation} 
which coincides\footnote{Note that the Fisher matrix that must be substituted into Eq.\ \eqref{eq:cramerfull} is still $-\langle \partial_j \partial_k p(s|\theta) \rangle_n = (h_j,h_k)$, and not $-\langle \partial_j \partial_k [p(s|\theta) p(\theta)] \rangle_n = (h_j|h_k) + P_{jk}$. The prior distribution does not concern the Cram\'er--Rao bound, which is computed from the likelihood alone for a \emph{fixed known} value of the true source parameters. Instead, we happen to be using an estimator that takes into in account prior information, which enters into the Cram\'er--Rao bound via the derivative of the bias.} with the generalized Cram \'er--Rao bound of Eq.\ \eqref{eq:cramerfull}, proving that the estimation problem defined by the LSA likelihood and $\hat{\theta}^\mathrm{MP}$ belongs to the exponential family.

The reason why $\hat{\theta}^\mathrm{MP}$ is not too useful to characterize future parameter-estimation performance is that we expect a reasonable measure of error to converge to the effective width of the prior in the limit of vanishing signal strength. Instead,
in the absence of any information from the experiment, $\hat{\theta}^\mathrm{MP}$ becomes stuck at the mode of the prior, and its variance [in Eq.\ \eqref{eq:varwithprior}] tends to zero. This behavior occurs for any nonuniform prior.\footnote{For uniform priors (e.g., rectangular distributions corresponding to the allowed parameter ranges), $\hat{\theta}^\mathrm{MP}$ actually becomes undefined in the $A \rightarrow 0$ limit.}

\subsection{Third-road Derivation of the Bayesian high-SNR/LSA result}
\label{sec:bayeshighsn}

We now wish to show that in any single experiment, if the high-SNR/LSA limit is warranted (and if the parameter priors are uniform over the parameter region of interest), the inverse Fisher-information matrix yields the variance of the Bayesian posterior probability distribution. To do so, we rewrite Eq.\ \eqref{eq:likexpanded} in terms of \emph{normalized parameters} $\bar{\theta}_i = A \, \theta_i$:
\begin{multline}
\label{eq:anotherp}
p(s|\theta) \propto \exp \Bigl\{ -(n,n)/2 + \Bigl[ (n,\bar{h}_j) \bar{\theta}_j + 
\frac{1}{A} (n,\bar{h}_{jk}) \bar{\theta}_{j} \bar{\theta}_{k} / 2 + 
\frac{1}{A^2} (n,\bar{h}_{jkl}) \bar{\theta}_{j} \bar{\theta}_{k} \bar{\theta}_{l} / 3! + O(1/A^3) \Bigr] \\
- \Bigl[ (\bar{h}_j,\bar{h}_k) \bar{\theta}_j \bar{\theta}_k +
\frac{1}{A} (\bar{h}_j,\bar{h}_{kl}) \bar{\theta}_j \bar{\theta}_{k} \bar{\theta}_{l}
+ \frac{1}{A^2} (\bar{h}_{jk},\bar{h}_{lm}) \bar{\theta}_{j} \bar{\theta}_{k} \bar{\theta}_{l} \bar{\theta}_{m} / 4 +
\frac{2}{A^2} (\bar{h}_{j},\bar{h}_{klm}) \bar{\theta}_{j} \bar{\theta}_{k} \bar{\theta}_{l} \bar{\theta}_{m} / 3! + O(1/A^3) \Bigr]/2
\Bigr\}.
\end{multline}
We can build the variance from the posterior mean
\begin{equation}
\label{eq:bayesianmean}
\bigl< \bar{\theta}_i \bigr>_p \equiv
\int \bar{\theta}_i \, p(s|\theta) \, d\theta
\bigg/\!\! \int p(s|\theta) \, d\theta
\end{equation}
and the quadratic moment
\begin{equation}
\bigl< \bar{\theta}_i \bar{\theta}_j \bigr>_p = 
\int \bar{\theta}_i \bar{\theta}_j \, p(s|\theta) \, d\theta
\bigg/\!\! \int p(s|\theta) \, d\theta
\end{equation} 
where ``$\langle \cdot \rangle_p$'' denotes integration over $p(s|\theta)$. The idea is to proceed in perturbative fashion, writing the moments as series in $\epsilon = 1/A$: taking $\langle \bar{\theta}_{i} \rangle_{p}$ as an example,
\begin{equation}
\big\langle \bar{\theta}_{i} \big\rangle_{p} =
\sum_{n=0}^\infty \frac{\epsilon^n}{n!}
\big\langle \bar{\theta}_{i} \big\rangle_{p}^{\!(n)}
\quad \Rightarrow \quad
\langle \bar{\theta}_i \rangle_p^{(n)} = \frac{\partial^n \langle \bar{\theta}_i \rangle_p}{\partial \epsilon^n} \bigg|_{\epsilon = 0}.
\end{equation}
Since $\epsilon$ appears at both the numerator and denominator of Eq.\ \eqref{eq:bayesianmean}, we write
\begin{equation}
\big\langle \bar{\theta}_{i} \big\rangle_{p} =
\frac{
\int \bar{\theta}_{i} \, p(0) \, d \theta +
\epsilon \int \bar{\theta}_{i} \, \frac{\partial p(0)}{\partial \epsilon} \, d \theta
+ \frac{\epsilon^2}{2} \int \bar{\theta}_{i} \, \frac{\partial^2 p(0)}{\partial \epsilon^2}  d \theta + \cdots
}{
\int p(0) \, d \theta +
\epsilon \int \frac{\partial p(0)}{\partial \epsilon} \, d \theta
+ \frac{\epsilon^2}{2} \int \frac{\partial^2 p(0)}{\partial \epsilon^2} d \theta + \cdots
}
\end{equation}
(where the argument of $p$ implies that the $(n)$-th derivative is evaluated at $\epsilon = 0$), and therefore
\begin{equation}
\begin{aligned}
\big\langle \bar{\theta}_{i} \big\rangle_{p}^{(0)} &=
\int \bar{\theta}_{i} \, p(0) \, d \theta \bigg/\!\!
\int p(0) \, d \theta,
\\
\big\langle \bar{\theta}_{i} \big\rangle_{p}^{(1)} &=
\biggl[ \int \bar{\theta}_{i} \, \frac{\partial p(0)}{\partial \epsilon} \, d \theta
- \big\langle \bar{\theta}_{i} \big\rangle_{p}^{(0)}
\int \frac{\partial p(0)}{\partial \epsilon} \, d \theta \biggr] \bigg/\!\! \int p(0) \, d \theta;
\\
& \ldots
\end{aligned}
\label{eq:bayesianseries}
\end{equation}
similar expressions hold for $\langle \bar{\theta}_i \bar{\theta}_j \rangle_p$, and a general expression for the $(n)$-th--order contribution is given in Sec.\ \ref{sec:higherbayesian}.
The $\epsilon \rightarrow 0$ limit coincides with the limit of large signal strengths, or of vanishing derivatives higher than the first, since in that case Eq.\ \eqref{eq:anotherp} truncates to Eq.\ \eqref{eq:standardlikelihood}. In this limit, 
\begin{equation}
\label{eq:bayesavg}
\bigl< \bar{\theta}_i \bigr>_p = \bigl< \bar{\theta}_i \bigr>_p^{(0)} = (\bar{h}_i,\bar{h}_j)^{-1} (n,\bar{h}_j)
\quad \text{(LSA)}
\end{equation}
and
\begin{equation}
\label{eq:bayesvar}
\bigl\langle \Delta \bar{\theta}_i \Delta \bar{\theta}_j \bigr\rangle_p = \bigl\langle (\bar{\theta}_i - \bigl\langle \bar{\theta}_i \bigr\rangle_p^{(0)}) (\bar{\theta}_j - \bigl\langle \bar{\theta}_j \bigr\rangle_p^{(0)}) \bigr\rangle_p^{(0)} = (\bar{h}_i,\bar{h}_j)^{-1}
\quad \text{(LSA)},
\end{equation}
and therefore
\begin{equation}
\bigl\langle \theta_i \theta_j \bigr\rangle_p = \frac{1}{A^2} (\bar{h}_i,\bar{h}_j)^{-1} = (h_i,h_j)^{-1} \quad \text{(LSA)},
\end{equation}
as can be seen by rewriting the exponential of Eq.\ \eqref{eq:standardlikelihood} as
\begin{equation}
p(s|\bar{\theta}) \propto
\exp \bigl\{ - (\bar{h}_i,\bar{h}_j) (\bar{\theta}_i - \langle \bar{\theta}_i \rangle_p) (\bar{\theta}_j - \langle \bar{\theta}_j \rangle_p) / 2 \bigr\},
\end{equation}
where we have omitted factors independent from $\bar{\theta}$ that cancel out in the normalization of $p(s|\bar{\theta})$.

Reinstating $A$ in Eq.\ \eqref{eq:bayesavg}
we see that in the high-SNR/LSA limit the mean of the posterior distribution coincides with the ML estimator, as is reasonable, since the average of a normal distribution coincides with its mode. The two however differ when higher-order terms are included, as we shall see in Sec.\ \ref{sec:higherorder}. From Eq.\ \eqref{eq:bayesvar} we see also that, to leading order, the variance of the posterior distribution is experiment-independent, and it coincides with the variance of the ML estimator (remember however that the two have very different interpretations\footnote{If we define the quadratic error of the posterior distribution as $\bigl< \bar{\theta}_i \bar{\theta}_j \bigr>_p$ (which is appropriate given that the true signal is at $\theta = 0$), we must increment $(\bar{h}_i,\bar{h}_j)^{-1}$ by the experiment-dependent quantity
$\bigl< \bar{\theta}_i \bigr> \bigl< \bar{\theta}_j \bigr> = (\bar{h}_i,\bar{h}_l)^{-1} (n,\bar{h}_l) 
(\bar{h}_m,n) (\bar{h}_m,\bar{h}_j)^{-1}$. Interestingly, the frequentist average of the Bayesian error $\bigl< \bigl< \bar{\theta}_i \bar{\theta}_j \bigr>_p \bigr>_n$ is $2 (\bar{h}_i,\bar{h}_j)^{-1}$, twice the frequentist variance of $\hat{\theta}^\mathrm{ML}$.}).

With the addition of a Gaussian prior $p(\theta) \propto e^{-P_{ij} \theta_i \theta_j / 2}$ centered at $\theta = 0$, Eqs.\ \eqref{eq:bayesavg} and \eqref{eq:bayesvar} change only slightly:\footnote{With the Gaussian prior, the quadratic error $\bigl< \bigl< \bar{\theta}_i \bar{\theta}_j \bigr>_p \bigr>_n$ becomes $[(\bar{h}_i,\bar{h}_l) + P_{il}/A^2]^{-1} (\bar{h}_l,\bar{h}_m) [(\bar{h}_m,\bar{h}_j) + P_{mj}/A^2]^{-1}$.}
\begin{equation}
\label{eq:resummedprior}
\begin{aligned}
\bigl\langle \bar{\theta}_i \bigr\rangle_p &= [(\bar{h}_i,\bar{h}_j) + P_{ij}/A^2]^{-1} (n,\bar{h}_j) \\
\bigl\langle \Delta \bar{\theta}_i \Delta \bar{\theta}_j \bigr\rangle_p &= [(\bar{h}_i,\bar{h}_j) + P_{ij}/A^2]^{-1}
\end{aligned} \quad \text{(LSA/Gaussian prior)}.
\end{equation}
Note that $p(\theta) \propto e^{-(1/A^2) P_{ij} \bar{\theta}_i \bar{\theta}_j/2}$ is formally an $O(1/A^2)$ contribution to the likelihood exponential that would enter the $1/A$ expansion beginning at that order. However, if $P_{ij}$ is large enough to matter at the signal strengths of interest, it probably makes sense to bundle it with the zeroth-order terms as we did here. In contrast with Eq.\ \eqref{eq:varwithprior} for the frequentist variance of $\hat{\theta}^\mathrm{MP}$, we see that in the limit of vanishing signal strength the variance of the posterior goes to the variance $P_{ij}$ of the prior.

\section{Standard compact-binary signal model}
\label{sec:standardsignal}

Throughout the rest of this paper, our fiducial model for compact-binary signals will be simple stationary-phase--approximated (SPA) waveforms including phasing terms from the spin--orbit and spin--spin interactions of parallel or antiparallel component spins. Parameter estimation with these waveforms was studied by Poisson and Will \cite{pw1995}. In this paper we adopt second-order post-Newtonian\footnote{Waveform phasing expressions accurate to 3.5PN order are also provided in Ref.\ \cite{arun2005}. We do not use these in this article for the sake of simplicity, since they would not change the qualitative picture of parameter estimation presented here. For the reader's reference, however, the higher-than-2PN corrections to Eq.\ \eqref{eq:pnphasing}, including the errata to Ref.\ \cite{arun2005}, are
\begin{multline}
\pi \biggl(\frac{38645}{252} - \frac{65}{9}\eta^3\biggr) v^5 \log v +
\biggl[ \biggl(\frac{11 583 231 236 531}{4 694 215 680} - \frac{640}{3} \pi^2 - \frac{6 848}{21} \gamma - \frac{6 848}{21}\log(4) \biggr)
+       \biggl(-\frac{15 335 597 827}{3 048 192} + \frac{2 255}{12}\pi^2  \\ - \frac{1 760}{3} \theta + \frac{12 320}{9} \lambda\biggr) \eta + \frac{76 055}{1 728}\eta^2 - \frac{127 825}{1 296}\eta^3  \biggr] v^6
-\frac{6 848}{21} v^6 \log v
+ \pi \biggl(\frac{77 096 675}{254 016} + \frac{378 515}{1 512}\eta - \frac{74 045}{756}\eta^2\biggr) v^7, \nonumber
\end{multline}
where $\gamma = 0.57721\cdots$ is Euler's constant, and $\lambda = -1987/3080$ and $\theta = -11831/9240$ are recently determined constants in the PN expansion \cite{blanchet2004}.} (2PN) Fourier-domain waveforms as written by Arun and colleagues \cite{arun2005}:
\begin{equation}
\label{eq:pnwaveform}
\tilde{h}(M_c,\eta,\beta,\sigma,\phi_0,t_0;f) \propto f^{-7/6} \exp \, i \{ \psi(M_c,\eta,\beta,\sigma;f) + \phi_0 + 2 \pi f t_0 \},
\end{equation}
with
\begin{equation}
\label{eq:pnphasing}
\begin{aligned}
\psi(M_c,\eta,\beta,\sigma;f) = \frac{3}{128 \, \eta \, v^5} \biggl\{
1 & +
\frac{20}{9} \biggl(\frac{743}{336} + \frac{11}{4} \eta\biggr) v^2 +
(4 \beta - 16 \pi) v^3 +
10 \biggl(\frac{3 058 673}{1 016 064} + \frac{5 429}{1 008}\eta + \frac{617}{144}\eta^2 - \sigma \biggr) v^4 \biggr\},
\end{aligned}
\end{equation}
where $v = (\pi M f)^{1/3}$, $M = m_1 + m_2$ is the total mass of the binary, $\eta = m_1 m_2 / M^2$ is the symmetric mass ratio, $M_c = M \eta^{3/5}$ is the chirp mass. The spin--orbit parameter $\beta$ and the spin--spin parameter $\sigma$ \cite{pw1995,spinpapers} are given by
\begin{equation}
\beta = \sum_{i=1}^{2} \frac{\hat{\mathbf{L}} \cdot \mathbf{S}_i}{12 \, m_i^2} \biggl[ 113 \Bigl(\frac{m_i}{M}\Bigr)^2 + 75 \, \eta \biggr] = \frac{113 \, \hat{\mathbf{L}} \cdot [\mathbf{S}_1 + \mathbf{S}_2] + 75 \, \hat{\mathbf{L}} \cdot [(m_2/m_1) \mathbf{S}_1 + (m_1/m_2) \mathbf{S}_2]}{12 \, M^2},
\end{equation}
and
\begin{equation}
\sigma = \frac{721 (\hat{\mathbf{L}}\cdot\mathbf{S}_1)(\hat{\mathbf{L}}\cdot\mathbf{S}_2) - 247 \,\mathbf{S}_1 \cdot \mathbf{S}_2}{48 \, m_1 \, m_2 \, M^2},
\end{equation}
with $\mathbf{S}_1$ and $\mathbf{S}_2$ the spins of the binary components. We truncate waveforms at the (Keplerian) last stable circular orbit ($v = 1/\sqrt{6}$).

For simplicity, in this article we do not discuss the estimation of the amplitude parameter $\mathcal{A}$ that would multiply the right-hand side of Eq.\ \eqref{eq:pnwaveform}. [From Eqs.\ \eqref{eq:noiseproduct} and \eqref{eq:pnwaveform} it follows that $(\partial_\mathcal{A} h,\partial_i h) = 0$ for $i \neq \mathcal{A}$, so the amplitude $\mathcal{A}$ effectively decouples from all other parameters in the Fisher matrix.] However, all discussions to follow can accommodate the addition of $\mathcal{A}$ with trivial modifications. 

\section{The singular case of the disappearing parameter}
\label{sec:disappearing}

In Sec.\ \ref{sec:threeways} we have examined the interpretation of the inverse Fisher matrix as a frequentist or Bayesian measure of error or uncertainty.
In this section, we discuss what happens when the Fisher is matrix singular, or almost so, so that the attempts to invert it numerically yield warnings that it is badly conditioned. It is pedagogical to begin this discussion by considering the case where the matrix is exactly singular (Sec.\ \ref{sec:singularfisher}), and then to widen our scope to approximate singularity (Sec.\ \ref{sec:approxfisher}). The conclusion is that a singular Fisher matrix is almost always a symptom that the high-SNR/LSA limit is not to be trusted, that prior probabilities play an important role, or both. 

\subsection{Singular Fisher matrix}
\label{sec:singularfisher}

A singular Fisher matrix implies that the corresponding LSA likelihood \eqref{eq:standardlikelihood} is a \emph{singular normal} distribution \cite{gupta},
which is constant along the directions of the Fisher-matrix eigenvectors with null eigenvalues\footnote{\label{note:coordinates}A reasonable objection to computing the eigensystem of the Fisher matrix is that it leads to taking linear combinations of parameters that may have different units. It is possible to avoid this problem by looking at the Fisher matrix more abstractly as a linear operator, and talking of its range and null space \cite{tarantola}; or more pragmatically, by dividing all parameters by their typical range; or perhaps by taking their logarithm (since we are working with errors, units can be forgotten as additive constants), which in the linearized theory is equivalent to dividing by the true parameters. We are going to largely ignore this issue, treating the parameters as pure numbers resulting from adopting a God-given system of units.} (henceforth, somewhat improperly, we shall call these \emph{null eigenvectors}), so
the ML equation has no solutions, and the even moments of the distribution are infinite, even for parameters that do not appear in the null eigenvectors. Thus, the frequentist variance of the ML estimator and the Bayesian variance of the posterior distribution are (formally) infinite for all parameters.

How to deal with this? If the signal is \emph{really} linear, so that the LSA expressions are exact, it is possible to discard the combinations of parameters that correspond to the null eigenvectors, and characterize the variance of the remaining parameters. Let us see how, in the frequentist and Bayesian frameworks. In what follows, we denote the total number of source parameters by $N$, and the number of non-null eigenvectors by $R$.

In the frequentist ML framework, we write $(\bar{h}_i,\bar{h}_j)$ in the singular-value (SV) decomposition\footnote{For square matrices, the SV decomposition is essentially equivalent to an eigenvector--diagonal-matrix decomposition where we drop the rows and columns corresponding to the null eigenvalues and eigenvectors.} \cite{golub} as
$\sum_{\lambda^{(k)} \neq 0} \theta^{(k)}_i \lambda^{(k)} \theta^{(k)}_j$ (with $(k) = 1,\ldots,R$), or $\Theta \Sigma \Theta^T$ in matrix notation (with $\Theta$ an $N \times R$ matrix with orthonormal columns, and $\Sigma$ a diagonal matrix formed from the $R$ non-zero eigenvalues).
We can then refactor the ML equation as
\begin{equation}
\label{eq:svmoment}
\bigl(\Theta \Sigma \Theta^T\bigr) \boldsymbol{\hat{\theta}} = A^{-1} \mathbf{\hat{n}}
\quad \Rightarrow \quad
\bigl(\Theta^T \boldsymbol{\hat{\theta}}\bigr) = A^{-1}
\bigl(\Sigma^{-1} \Theta^T \mathbf{\hat{n}}\bigr)
\quad \Rightarrow \quad
\hat{c}^{(k)} = A^{-1} (\lambda^{(k)})^{-1} n^{(k)},
\end{equation}
where $\hat{c}^{(k)}$ and $n^{(k)}$ denote the coefficients of the decompositions of $\hat{\theta}$ and $(\bar{h}_i,n)$ with respect to the normalized non-null eigenvectors of $(\bar{h}_i,\bar{h}_j)$. Since the ensemble average $\langle n^{(k)} n^{(l)} \rangle_n$ is just $\lambda^{(k)} \delta^{(k)(l)}$ (where $\delta$ is Kronecker's delta), the frequentist covariance of the ML estimators $\hat{c}^{(k)}$ is the diagonal matrix $A^{-2} (\lambda^{(k)})^{-1} \delta^{(k)(l)}$.

In the Bayesian framework, the quantities of interest are the moments of the $c^{(k)}$ over infinite ranges of the $c^{(k)}$ \emph{and} of the coefficients $C^{(K)}$ (with $(K) = 1,\ldots,N-R$) corresponding to the null eigenvectors, which are not included in the SV decomposition. Formally, these moments are ratios of two infinities,
because the LSA likelihood is not a function of the $C$ [not even through the $n^{(K)} \equiv \theta_i^{(K)} (\bar{h}_i,n)$ terms, which are zero since $(\theta_i^{(K)} \bar{h}_i,\theta_j^{(K)} \bar{h}_j) = 0$],
but they may be evaluated as improper integrals, \emph{in the limit} of the ranges for the $c^{(K)}$ extending to infinity:
\begin{equation}
\label{eq:limitmoment}
\langle \Delta c^{(k)} \Delta c^{(l)} \rangle_p = \frac{\int c^{(k)} c^{(l)} p(s|c) \, dc \, dC}{\int p(s|c) \, dc \, dC} = \lim_{\Delta C^{(K)} \rightarrow \infty}
\frac{\left(\int_{-\Delta C^{(K)}}^{+\Delta C^{(K)}} dC \right) \int c^{(k)} c^{(l)} p(s|c) \, dc}{\left( \int_{-\Delta C^{(K)}}^{+\Delta C^{(K)}} dC \right) \int p(s|c) \, dc} = A^{-2} (\lambda^{(k)})^{-1} \delta^{(k)(l)}.
\end{equation}

We can then work back to the frequentist components of the covariance matrix (or the Bayesian posterior moments) 
that involve any $\hat{\theta}_i$ that \emph{do not appear} in the null eigenvectors. All other $\hat{\theta}_i$, however, are \emph{completely} indeterminate.\footnote{In a truly linear system, this is true no matter how small the eigenvector component in that parameter direction; clearly, this raises a problem of accuracy in the numerical computation of eigenvectors.}
In the frequentist framework, it may be possible to work back to \emph{interval} estimates of their values by combining a ML estimate of the $\hat{c}^{(k)}$ with finite allowed ranges for some of the $\theta_i$; however, this would constitute a form of prior distribution for the $\theta_i$, which is not entirely compatible with the ML estimator (what happens if the solution of the ML equation falls outside the allowed range?). In the Bayesian framework, salvation may come from the prior probability distributions that make the posterior integrable.\footnote{Even a single prior in the form of a rectangle function will regularize the integration over all the null-eigenvector coordinates that include that parameter. For normal priors, whether the posterior becomes integrable depends on the eigenstructure of $A^2 (\bar{h}_i,\bar{h}_j) + P_{ij}$.} Unless the priors are also normal, though, the resulting moments cannot be expressed simply as analytical expressions of the Fisher matrix.

The most benign outcome occurs when the null eigenvectors correspond individually to one or more of the original parameters, or when the subspace spanned by null eigenvectors corresponds to a subset of the original parameters. The null-eigenvector combinations of parameters may also have clear physical interpretations: for instance, for a monochromatic, continuous sinusoid of frequency $f$, the absolute time offset $t_0$ and the initial phase $\phi_0$ are essentially the same parameter, so the Fisher matrix has a null eigenvector along the parameter combination $f t_0 - \phi_0$, which can be discarded, while $f t_0 + \phi_0$ remains well determined. A similar case is the degeneracy between luminosity distance and a certain function of the sky-position angles in the analysis of short GW chirps with a single ground-based detector.\footnote{Although neither of these examples is a linear model described exactly by the LSA, the degeneracy persists in the exact likelihood, so its Fisher-matrix diagnosis is correct. For such ``perfect'' degeneracies to occur, the two parameters must appear in all waveform expressions only as a sum or product; this would imply that their units can be sensibly summed, or that their combination has direct physical meaning.} Other combinations of parameters can be more ambiguous and troubling---what is the meaning of estimating a parameter equal to a mass plus a spin? In those cases, our best hope is again that the degeneracy will be cured by prior probabilities, or by higher-order corrections in the $1/A$ expansion, in which cases the Fisher-matrix formalism is certainly insufficient.

\subsection{Ill-conditioned Fisher matrix}
\label{sec:approxfisher}

All nonsingular matrices have well-defined inverses, although these \emph{may} be difficult to compute. The notion of \emph{ill conditioning} from the theory of linear systems of equations
\cite{golub} can be invoked here to provide a bound (valid under reasonable conditions) on the perturbation of the inverse of a perturbed matrix,
\begin{equation}
\label{eq:conditionm}
\frac{||(M + \delta M)^{-1} - M^{-1}||}{||M^{-1}||} \leq
\kappa(M) \frac{||\delta M||}{||M||} + O(||\delta M||^2);
\end{equation}
here ``$||\cdot||$'' is a matrix norm (e.g., the 2-norm 
$||M||_2 = \sup_\mathbf{x} ||M \mathbf{x}||_2 / ||\mathbf{x}||_2$ derived from the vector 2-norm 
$||\mathbf{x}||_2 = (\sum_i x_i^2)^{1/2}$), and $\kappa(M) = ||M||\,||M^{-1}||$ is the \emph{condition number}. Since $||M||_2$ is equal to $M$'s largest eigenvalue, $\kappa_2(M)$ is given by the ratio of its largest- to smallest-modulus eigenvalues.
From a numerical-analysis perspective, as Finn \cite{finn1992} points out,
the gist of Eq.\ \eqref{eq:conditionm} is that, roughly speaking, matrix inversion can amplify roundoff error by a factor $\kappa$, leading to the loss of up to $\log_{10} \kappa$ digits of precision. The same amplification will apply to any inaccuracies in our knowledge of $M$. Taken at face value, this means that the Fisher-matrix results of Eqs.\ \eqref{eq:lsavar} and \eqref{eq:bayesvar} may be inaccurate at a 100\% level if the components of the Fisher matrix are not known to a fractional accuracy better than $\kappa^{-1}(F)$.

Of course, Eq.\ \eqref{eq:conditionm} is only an upper bound, and this doomsday scenario needs not be realized in practice. One way to check whether the matrix-inversion sensitivity is a concern is to add small random perturbations, Monte Carlo-style, to the Fisher matrix elements, and then verify the change in the covariance matrix. Such an experiment for our standard SPA model
(with $m_1 = m_2 = 10 M_\odot$ and no spins) shows that perturbing the \emph{12th} significant digits of the $F_{ij}$ components is already enough to engender 100\% changes in the diagonal elements of $(F^{-1})_{ij}$ (i.e., the predicted parameter variances).
This behavior is $\sim 100$ times less severe than what is predicted by Eq.\ \eqref{eq:conditionm},
but it still tells a rather cautionary tale about numerical sensitivity in the inversion of that particular Fisher matrix.\footnote{Augmenting the Fisher matrix with normal priors for $\eta$, $\beta$, and $\sigma$, as described in Sec.\ \ref{sec:priorexample}, can somewhat cure this instability to inversion, although the result is SNR-dependent: for $\mathrm{SNR}=10$, the errors in $(F_{ij} + P_{ij})^{-1}$ become intolerable for fractional perturbations in $F_{ij}$ of order $10^{-7}$ rather than $10^{-12}$, but in the high-SNR limit the threshold reverts to the latter.} These problems can be cured, somewhat trivially, by adopting higher-precision arithmetics, and by computing the Fisher matrix to better accuracy. It may also be possible to improve the condition number by changing the units of source parameters, which may reduce the magnitude gap between the largest- and smallest-modulus eigenvalues.

More to the point, it is the consequences of the Fisher-matrix condition number on the substance (rather than the numerical accuracy) of Eqs.\ \eqref{eq:lsaest} and \eqref{eq:bayesavg} that should attract our attention. A large condition number implies one or more small Fisher-matrix eigenvalues, and consequently large statistical fluctuations for the combinations of source parameters corresponding to the small-eigenvalue eigenvectors, at least according to the LSA. The interpretation is that large parameter changes in the direction of the small-eigenvalue eigenvectors are needed to produce changes in the waveform comparable to typical noise fluctuations. Under this condition, we have to worry whether the LSA can really describe the likelihood over the entire parameter ranges of interest: of course, these depend on the SNR available at detection (at leading order, their extent is inversely proportional to signal strength). In Sec.\ \ref{sec:howhigh} we describe a numerical criterion to decide when the SNR is high enough to believe the LSA. We also have to worry whether prior probability distributions for the parameters (perhaps in the simple form of allowed ranges) already restrict the estimated (for frequentists) or probable (for Bayesians) values of parameters beyond what is predicted by the Fisher-matrix variance. In the next section we discuss a simple test to decide whether priors should be included.

\section{The burden of prior commitments}
\label{sec:priors}

As Cutler and Flanagan point out \cite[p.\ 2691]{cf1994}, ``it is not necessary for \emph{a priori} information to be very detailed or restrictive in order that it have a significant effect on parameter-extraction accuracy. All that is necessary is that it be more restrictive than the information contained in the waveform, for some of the parameters [\ldots] what is more surprising is that due to the effects of correlations, the rms errors obtained for the other parameters may also be overestimated by large factors.'' Roughly speaking, this happens because as we move in parameter space, the change in the signal can be partially absorbed by changing correlated parameters together; thus, limiting the range available to one parameter also limits the range over which a correlated parameter can run while not significantly modifying the signal. In this section we seek a practical recipe to determine, in the context of a parameter-estimation problem specified by a family of waveforms and a fiducial SNR, whether it is necessary to take priors into consideration when evaluating projected parameter accuracies.

Since prior probabilities can only be discussed consistently in the framework of Bayesian parameter estimation, in this section we will restrict ourselves to that context. The Gaussian priors examined at the end of Sec.\ \ref{sec:bayeshighsn} are rarely appropriate in actual practice, but they do provide a quick test to see if the prior-less Fisher result can be taken as it stands, or whether a more careful analysis is needed that includes the effects of priors. In Sec.\ \ref{sec:priorexample} we try out this quick test on the SPA model of Sec.\ \ref{sec:standardsignal}. For simplicity, we shall consider the effects of priors as logically independent from the sufficiency of the LSA, although the two problems clearly come into play together in real situations.

\subsection{Testing for the influence of priors (normal true-parameter--centered priors)}
\label{sec:priorexample}

We shall discuss our quick test by way of an example.
The standard SPA model of Sec.\ \ref{sec:standardsignal} has six parameters: $M_c$, $\eta$, $\beta$, $\sigma$, $\phi_0$, and $t_0$ (plus $A$, which we disregard). We work at 2PN with SNR = $A$ = 10, with true parameters $m_1 = m_2 = 10 M_\odot$ (corresponding to $M_c = 8.71 M_\odot$, $\eta = 0.25$), and $\beta = \sigma = \phi_0 = t_0 = 0$. We wish to examine the effect of priors for three related parameter-estimation problems involving different subsets of parameters:
a 4-parameter problem (4pp) where we disregard spin parameters (i.e., where we assume we know \emph{a priori} that the true binary has no spin); a 5pp where spin--orbit coupling 
[as represented by $\beta$ in Eq.\ \eqref{eq:pnphasing}] is important, but spin--spin interactions can be neglected; and a 6pp where we include also spin--spin interactions [as represented by $\sigma$ in Eq.\ \eqref{eq:pnphasing}]. As we shall see, priors become increasingly important as the number of parameters increases.

In each problem, we compute the expected covariance matrix of the posterior distribution as the inverse of (a submatrix of) the Fisher matrix, neglecting any non-LSA effects. We represent priors as normal distributions centered around null parameter displacements (i.e., the true parameter value), with standard deviations of 0.25 for $\eta$ and, following Poisson and Will \cite{pw1995} 8.5 for $\beta$ and 5 for $\sigma$ (in Ref.\ \cite{baw2005}, Berti and colleagues derive and adopt approximate priors for $\beta$ and $\sigma$ with standard deviations $\Delta \beta = 9.4$ and $\Delta \sigma = 2.5$). This representation is very crude, but it is the only one that leads to a simple analytical result [Eq.\ \eqref{eq:resummedprior}] for the  posterior covariance, and it should give at least a qualitative idea of the effect of imposing rectangular priors covering the allowed parameter ranges. Results are shown in the upper section of Table \ref{tab:priors1010}, and are as follows.
%
%
\begin{table}
\begin{tabular}{llllllll}
& $\Delta M_c / M_c$ & \multicolumn{1}{c}{$\Delta \eta$} & \multicolumn{1}{c}{$\Delta \beta$} & $\Delta \sigma$ & $\Delta \phi_0$ & \multicolumn{2}{c}{$
\Delta t_0\,(\mathrm{ms})$} \\
\hline
4pp, no priors                                 & $2.9\times 10^{-2}$ & $8.3\times 10^{-2}$ &                  &     & 7.3 &\;\;& 3.0 \\
4pp, NTC prior on $\eta$                       & $2.7\times 10^{-2}$ & $7.9\times 10^{-2}$ &                  &     & 7.0 && 2.8 \\
5pp, no priors                                 & \underline{$1.1$}               & \underline{$5.1\times 10^1$}
                                                                                           & \underline{$1.2 \times 10^3$} &   & $3.3 \times 10^2$ && 6.9 \\
5pp, NTC prior on $\beta$                      & $3.0\times 10^{-2}$ & \underline{$3.8\times 10^{-1}$} & 8.5              &     & 7.7 && 3.0 \\
5pp, NTC prior on $\eta$, $\beta$              & $3.0\times 10^{-2}$ & $2.1\times 10^{-1}$ & 4.9              &     & 7.7 && 3.0 \\
6pp, NTC priors on $\eta$, $\beta$             & $4.3\times 10^{-2}$ & $2.5\times 10^{-1}$ & 8.4  & \underline{$2.5 \times 10^1$}
                                                                                                                    & $4.3 \times 10^1$ && 5.3 \\
6pp, NTC priors on $\eta$, $\beta$, $\sigma$   & $3.0\times 10^{-2}$ & $2.1\times 10^{-1}$ & 5.1              & 4.9 & $1.1 \times 10^1$ && 3.1 \\
\hline
4pp, exact priors on $\eta$          & $1.8\times 10^{-2}$ & $5.0\times 10^{-2}$ &                  &     & 4.4 && 1.9 \\
5pp, exact priors on $\eta$, $\beta$ & $2.9\times 10^{-2}$ & $7.1\times 10^{-2}$ & 2.4              &     & 7.5 && 2.9 \\
6pp, exact priors on $\eta$, $\beta$, $\sigma$ & $2.9\times 10^{-2}$ & $7.1\times 10^{-2}$ & 2.6    & 2.9 & 9.0 && 3.0 \\
\hline
\end{tabular}
\caption{
Fisher-matrix rms errors in the 4-, 5-, and 6-parameter--estimation problems for a $(10+10)M_\odot$ binary with $\beta = \sigma = 0$ and $\mathrm{SNR} = 10$, evaluated under different combinations of normal true-parameter--centered (NTC) priors (upper section of table) and of the exact priors of Sec.\ \ref{sec:exactprior} (lower section). The underlined errors are larger than the physical range for the parameter.
\label{tab:priors1010}}
\end{table}

The first line of Table \ref{tab:priors1010} shows the 4pp no-prior $1\sigma$ values for the single-parameter rms errors (i.e., the square roots of the diagonal elements in the covariance matrix). Among these, $\Delta M_c$ and $\Delta \eta$ seem reasonable, but we get hung up on the value of $\Delta \phi_0$. Can the $1\sigma$ error region be larger than the physically meaningful range for this angle? On general grounds, we should worry that the LSA cannot know that the waveforms are exactly periodic (and therefore nonlinear) in the angular parameters, so it blithely extrapolates small-angle effects to infinite ranges. However, as pointed out by Cutler \cite{cpc2007}, this extrapolation is roughly correct for a simple complex phase such as $\phi_0$ [see Eq.\ \eqref{eq:pnwaveform}], for which the main correlated-parameter effect is to absorb the global phase shifts due to changes in the other parameters.\footnote{In particular, $\phi_0$ is strongly coupled to $t_0$, which produces frequency-dependent phase shifts through the exponential $\exp(2 \pi i f t_0)$. Adopting the new phase parameter $\phi'_0 = \phi_0 + 2 \pi f_0 t_0$, where $f_0$ is the \emph{dominant} frequency at which $f^{-7/3} / S_n(f)$ is maximum, largely removes this coupling \cite{cpc2007}.} A large $\Delta \phi_0$ indicates that this absorption can happen through several cycles of phasing.  We conclude that $\phi_0$ is essentially undetermined, but we have no reason to distrust the errors for $\Delta M_c$, $\Delta \eta$, and $\Delta t_0$.

Applying a prior to $\eta$ does not change the picture significantly, but priors do matter once we add the spin parameters, which are very poorly determined at this SNR. In the 5pp, we find unphysically large errors for both $\beta$ \emph{and} $\eta$, which are cured only by imposing priors on \emph{both} parameters. In the 6pp, we find that a prior is needed also for $\sigma$; adding it engenders measurable changes in $\Delta M_c$ and $\Delta \eta$.  As a rule of thumb, we should expect such effects whenever the signal derivatives show significant correlations, and when the magnitudes of the priors, measured crudely as the squared inverses $(\theta_i^\mathrm{max} - \theta_i^\mathrm{min})^{-2}$ of the effective parameter ranges induced by the priors, are comparable to the corresponding diagonal Fisher-matrix elements $F_{ii}$.

\subsection{Testing for the influence of priors (exact priors)}
\label{sec:exactprior}

We can perform an even better test by evaluating the effects of \emph{exact} priors while still working in the LSA.
Doing this requires some numerics, which are however very manageable on a workstation-class system.
The idea is to integrate $\langle \Delta \theta_i \Delta \theta_j \rangle_p$ as a Monte Carlo sum, which can be accomplished as follows.
First, we need to fix a reference experiment by drawing the random variable $n_j \equiv (n,h_j)$ from its ensemble distribution, which in Gaussian noise is a normal distribution with mean zero and covariance matrix $F_{ij} \equiv (h_i,h_j)$ [see Eq.\ \eqref{eq:noiseprod}]. To do so, we generate a zero-mean, unit-variance, normal $N$-tuple, and multiply it by $\sqrt{F_{ij}}$ (where the square root is taken in the linear-operator sense and exists for nonsingular Fisher matrices). Note that we cannot work with SNR-invariant expressions (e.g., normalized parameter errors $\bar{\theta}_i$), since the priors set a scale for the strength of the signal.

We can now draw samples distributed with the LSA likelihood $p(s|\theta) \propto \exp \{ -(h_j,h_k) \theta_j \theta_l / 2 + n_j \theta_j \}$.
To do so, we generate zero-mean, unit-variance, normal $N$-tuples, multiply them by $(F_{ij}^{-1})^{1/2}$, and offset them by $F_{ij}^{-1} n^j$.
We include the effects of priors (therefore obtaining a population $\{\theta^{(i)}\}$ distributed according to the LSA \emph{posterior} probability) by going through the samples, and discarding each of them with a probability $1 - p(\theta) / \max_\theta p$ (for rectangular priors the probability of discarding is always 0 or 1). The covariance matrix of the posterior distribution can then be computed from the surviving samples.

We repeat this procedure for many different experiments (i.e., $n_j$'s), and take a frequentist average of the covariance-matrix components (or study their frequentist distribution). Again, the Bayesian interpretation of this entire procedure is as follows: we place the true signal at $\theta = 0$; we draw from the possible noise realizations according to their ensemble probability; and we compute the variance of the posterior distribution for each noise realization.
If the priors are very restrictive compared to Fisher-matrix--only errors, we may find that we are discarding a very large percentage of the samples. To avoid this, we can incorporate a normal approximation to the priors in the probability distribution used to generate the samples (i.e., by multiplying normal variates by $\sqrt{[(h_i,h_j) + P_{ij}]^{-1}}$, and offsetting them by $[(h_i,h_j) + P_{ij}]^{-1} n^j$), and then sieve the resulting samples with respect to $\propto p(\theta) e^{P_{ij} \theta_i \theta_j / 2}$ instead of $p(\theta)$. It is also possible to use
rejection sampling \cite{nrc}, as we did for the results reported in this section, or
the Metropolis algorithm \cite{metropolis} with the likelihood or likelihood-plus-NTC-prior as proposal distribution, and the full posterior as the target distribution.

Applying the procedure outlined above to our $(10+10)M_\odot$ system yields the results listed in the lower section of Table \ref{tab:priors1010}. We adopt exact priors given by rectangular probability distributions covering the intervals $[0,\infty]$ for $M_c$, $[0,0.25]$ for $\eta$, $[-8.5,8.5]$ for $\beta$, and $[-5,5]$ for $\sigma$. Each quoted error is a frequentist average of 200 independent Monte Carlo estimates, each computed for a different realization of noise from an initial sampling of $10^6$ parameter sets, reduced to $5\times 10^4$--$2 \times 10^5$ samples after rejection sampling, depending on the estimation problem.

The expected errors are significantly reduced compared to the NTC-prior estimates. These reductions stem mainly from the greater tightness of the rectangular priors, and are especially significant for for $\eta$, for which the symmetric NTC prior is indeed very crude.  The lesson is that we can use $[(h_i,h_j) + P_{ij}]^{-1}$ (i.e., the quick test) to decide whether priors are important, but we need something more sophisticated, such as the procedure described in this section, to gauge their effects accurately. Of course, this gain in accuracy may be only virtual if the LSA is not warranted for our problem. Deciding that question is the object of the next section.

\section{The unbearable lightness of signal to noise}
\label{sec:howhigh}

As we have seen in Sec.\ \ref{sec:threeways}, the high-SNR and LSA limits coincide because larger signal strengths correspond to smaller statistical errors, which in turn imply that the linearized-signal expression \eqref{eq:standardlikelihood} for the likelihood is more accurate. The equivalence of the two limits is manifest in the $1/A$ expansions such as Eqs.\ \eqref{eq:expandsystem} and \eqref{eq:anotherp}.
Indeed, Finn \cite{finn1992} cautions that ``it is important that the probability contours of interest (e.g. 90\%) do not involve [errors] so large that the linearization of [the likelihood] is a poor approximation.''

In practice, given a family of waveforms and the true parameter values, we need to ask how high an SNR is needed for the limits to yield accurate expected errors. One approach involves comparing the Fisher-matrix results with errors computed at the next order in the $1/A$ expansions: in App.\ A 5 of Ref.\ \cite{cf1994}, Cutler and Flanagan provide next-order formulas for the frequentist variance (although they do not apply them to the Fisher-matrix estimates in the same article).
In Sec.\ \ref{sec:higherorder} we provide the mathematicals tools to do so in our notation; we must however warn the reader that the calculation is rather cumbersome, except for simple waveforms, and the verdict is still not definitive: the smallness of a term in a series does not guarantee that the series is converging.

A simpler approach involves working with the ratio $r(\theta,A)$ of the LSA likelihood to the exact likelihood to build a consistency criterion for the Fisher-matrix formalism. In this section we shall see that under reasonable conditions, the ratio $r$ is given in logarithm by
\begin{equation}
\label{eq:sntest}
|\log r(\theta,A)| =
\bigl(\theta_j h_j - \Delta h(\theta) , \theta_k h_k - \Delta h(\theta) \bigr) / 2,
\end{equation}
where $\Delta h(\theta) = h(\theta) - h(0)$, $A$ is the signal strength, and $\theta$ is the error (in a sense to be made precise shortly). Since $h(\theta) = h(0) + \theta_i h_i + \cdots$, the product in Eq.\ \eqref{eq:sntest} represents the noise-weighted norm of the higher-than-linear contributions to $h(\theta)$, expanded around the true source parameters. The idea of the criterion is to choose an isoprobability surface (say, the $1\sigma$ surface), as predicted by the Fisher matrix, and then explore it to verify that the mismatch between the LSA and exact likelihoods is smaller than a fiducial value (say, $|\log r| < 0.1$), so that we can actually believe the LSA in predicting the $1\sigma$ surface to begin with.

We stress that this is just a criterion of \emph{consistency}. Even if the Fisher-matrix result is internally consistent, it may still be inaccurate; conversely, the structure of the ambiguity function across parameter space could conspire in such a way as to make the LSA results correct, although we have no reason to expect that in general.
In the rest of this section, we explain how the criterion comes about in the frequentist (Sec.\ \ref{sec:criterionfreq}) and Bayesian (Sec.\ \ref{sec:criterionbayes}) frameworks, and we show a concrete example of the criterion in use (Sec.\ \ref{sec:criterionpractical}).

\subsection{Frequentist justification of the maximum-mismatch criterion}
\label{sec:criterionfreq}

\newcommand{\thetas}{\theta^{1\sigma}}

As in Secs.\ \ref{sec:threeways} and \ref{sec:priors}, we assume that the detector output is $s = A \bar{h}_0 + n$. In the LSA, the ML estimator $\hat{\theta}^\mathrm{ML}$ is a normal variable with mean zero and covariance matrix $(1/A^2) (\bar{h}_j,\bar{h}_k)^{-1}$ [Eq.\ \eqref{eq:noiseprod}]. For a given signal amplitude $A$, let $\hat{\theta}^\mathrm{ML}$ take on the specific value $\thetas$ on its $1\sigma$ surface. From Eqs.\ \eqref{eq:pstheta}, \eqref{eq:standardlikelihood}, and \eqref{eq:sntest}, the mismatch ratio $r$ is given by  
\begin{equation}
r(\thetas)
= \exp \left\{-\bigl(s - A (\bar{h}_0 + \thetas_j \bar{h}_j),s - A (\bar{h}_0 + \thetas_k \bar{h}_k)\bigr)/2\right\} \Big/ \exp\left\{-\bigl(s - A \bar{h}(\thetas), s - A \bar{h}(\thetas) \bigr)/2\right\};
\end{equation}
writing $s$ out, we eliminate all instances of $\bar{h}_0$:
\begin{equation}
\log r(\thetas) = - A^2 \thetas_j \thetas_k \bigl( \bar{h}_j, \bar{h}_k \bigr) / 2
+ A^2 \bigl( \Delta \bar{h}(\thetas), \Delta \bar{h}(\thetas) \bigr) / 2
+ A \, \thetas_j \bigl( \bar{h}_j, n \bigr)
- A \bigl( \Delta \bar{h}(\thetas), n \bigr).
\end{equation}
The first two terms in the exponent can be computed given $\thetas$; not so the two products involving $n$. To obtain the first, we note that if $\hat{\theta}^\mathrm{ML} = \thetas$, then the noise must be such that $\thetas = (1/A) (\bar{h}_j,\bar{h}_k)^{-1} (\bar{h}_k,n)$ [Eq.\ \eqref{eq:lsaest}], so $(\bar{h}_j,n) = A (\bar{h}_j,\bar{h}_k) \thetas_k$. To obtain the second, we change our perspective slightly, and average $\log r(\thetas)$ over all noise realizations $n$ compatible with $\thetas$. This is how. Let $x_j \equiv (\bar{h}_j,n)$, $y \equiv ( \Delta \bar{h}(\thetas), n )$: separately, $x_j$ and $y$ are normal random variables with mean zero and covariances equal to $(\bar{h}_j,\bar{h}_k)$ and $( \Delta \bar{h}(\thetas), \Delta \bar{h}(\thetas) )$, respectively; taken together, they are jointly normal variables with covariance $(\bar{h}_j,\Delta \bar{h}(\thetas))$ [Eq.\ \eqref{eq:noiseprod} again]. 
We now know enough to build $p(x,y)$, from which we can derive the conditional probability $p(y|x)$ and compute the conditional mean of $y$, which (after the algebra of App.\ \ref{sec:conditional}) turns out to be $A \, \thetas_j (\bar{h}_j,\Delta \bar{h}(\thetas))$. Altogether, we find 
\begin{equation}
\begin{aligned}
\langle \log r(\thetas) \rangle_{n(\thetas)} &= A^2 \thetas_j \thetas_k \bigl( \bar{h}_j, \bar{h}_k \bigr) / 2
+ A^2 \bigl( \Delta \bar{h}(\thetas), \Delta \bar{h}(\thetas) \bigr) / 2
- A^2 \thetas_j \bigl( \bar{h}_j,\Delta \bar{h}(\thetas) \bigr) \\
&=  \bigl( \thetas_j h_j - \Delta h(\thetas), \thetas_k h_k - \Delta h(\thetas) \bigr) / 2,
\label{eq:frequentistresult}
\end{aligned}
\end{equation}
just as anticipated in Eq.\ \eqref{eq:sntest}. The signal strength $A$ enters Eq.\ \eqref{eq:frequentistresult} explicitly, but also implicitly through the parameter width of the Fisher-matrix $1\sigma$ surface. Thus Eq.\ \eqref{eq:sntest} can be solved for the $A$ that corresponds to $\thetas$ small enough to yield $r$ as close to unity as desired. Since to leading order $\thetas_j - \Delta h(\thetas) = A \bar{h}_{jk} \thetas_j \thetas_k$, and since to leading order $\thetas$ scales as $1/A$, we expect $\log r$ to scale as $1/A^2$ for large enough $A$.

In summary, the maximum-mismatch criterion is justified from a frequentist viewpoint as a constraint on the ratio $r$ at points on a constant--LSA-probability surface, averaged over all realizations of noise compatible with finding the ML estimator at those points.

\subsection{Bayesian justification of the maximum-mismatch criterion}
\label{sec:criterionbayes}

The justification of the maximum-mismatch criterion from a Bayesian viewpoint requires another slight change of perspective. Again we assume $s = A \bar{h}_0 + n$; this time, however, we expand the waveform not with respect to the true parameters ($\theta = 0$), but to the observed location $\hat{\theta}^\mathrm{ML}(n) \equiv \theta^n_0$ of maximum LSA likelihood for a given experiment. In the absence of priors, it is with respect to this location that the uncertainty of the posterior would be judged in a single experiment.
Thus we write $h(\theta) \equiv h(\theta^n_0 + \theta^n) \simeq h^n_0 + \theta^n_j h^n_j$, where $h^n_0 \equiv h(\theta^n_0)$: the ``${}^n$'' superscripts serve to remind us that the parameter displacements $\theta^n$ (and the waveform derivatives $h^n_j$) are evaluated from (at) $\theta^n_0$. We also write $\Delta h^n_0$ for $h_0 - h^n_0$.

The $1\sigma$ surface over which we are going to evaluate the ratio $r$ will be a surface of equiprobable true-signal locations, given the observed location $\theta^n_0$ of maximum LSA likelihood. In the LSA, the distribution over experiments of the true-signal location with respect to $\theta^n_0$ is again normal with covariance matrix $(h^n_i,h^n_j)^{-1}$. Thus we have $\theta^n = -\theta^n_0$, and the mismatch ratio $r$ is given by
\begin{equation}
r(\theta_0) = \exp \left\{ -\bigl( s - A(\bar{h}_0^n + \theta^n_j \bar{h}^n_j), s - A(\bar{h}_0^n + \theta^n_k \bar{h}^n_k) \bigr) / 2 \right\} \Big/
\exp \left\{ -\bigl( s - A \bar{h}_0, s - A \bar{h}_0 \bigr) \right\};
\end{equation}
writing $s$ out, the denominator reduces to $\exp -(n,n)/2$, and $h_0$ enters the numerator only through $\Delta h^n_0$:
\begin{equation}
\label{eq:tempbayesianr}
\log r(\theta_0) = - A^2 \theta^n_j \theta^n_k \bigl( \bar{h}^n_j, \bar{h}^n_k \bigr) / 2
- A^2 \bigl( \Delta \bar{h}^n_0, \Delta \bar{h}^n_0 \bigr) / 2
+ A^2 \theta^n_i \bigl( \bar{h}^n_i, \Delta \bar{h}^n_0 \bigr)
+ A \, \theta^n_i \bigl( \bar{h}^n_i, n \bigr)
- A \bigl(\Delta \bar{h}^n_0, n \bigr).
\end{equation}

Now, since the LSA likelihood can be written as $p(s|\theta^n) \propto \exp -(n + \Delta h_0^n - \theta^n_i h_i^n,n + \Delta h_0^n - \theta^n_j h_j^n) / 2$, the ML equation $\partial p/\partial \theta^n_i = 0$ at $\theta^n_i = 0$ implies $(h^n_i,n) = -(h^n_i,\Delta h^n_0)$.
We handle the last term of the equation by evaluating 
the conditional mean of $y^n \equiv -A (\Delta \bar{h}^n_0, n)$ given $x^n = (\bar{h}^n_j,n) = -A (\bar{h}^n_j,\Delta \bar{h}^n_0)$, producing $-A (\bar{h}_j^n,n) (\bar{h}^n_j,\bar{h}^n_k)^{-1} (\bar{h}^n_k, \Delta \bar{h}^n_0) = A (\bar{h}_j^n,n) (h^n_j,h^n_k)^{-1} (h^n_k, n)$ (again, see App.\ \ref{sec:conditional}).
We can then use Eq.\ \eqref{eq:lsaest} to replace $(h^n_j,h^n_k)^{-1} (h^n_k, n)$ with $\theta^n_{0j} = -\theta^n_j$ (working to leading order), so that the last two terms of Eq.\ \eqref{eq:tempbayesianr} end up canceling out:
\begin{equation}
\begin{aligned}
\log r(\theta_0) &= - A^2 \theta^n_j \theta^n_k \bigl( \bar{h}^n_j, \bar{h}^n_k \bigr) / 2
+ A^2 \theta^n_j (\bar{h}^n_j, \Delta \bar{h}^n_0)
- A^2 \bigl( \Delta \bar{h}^n_0, \Delta \bar{h}^n_0 \bigr) / 2 \\
&= -\bigl( \theta^n_j h^n_j - \Delta h(\theta^n), \theta^n_k h^n_k - \Delta h(\theta^n) \bigr) / 2.
\label{eq:bayesianresult}
\end{aligned}
\end{equation}
Again, this equation can be solved for the $A$ that corresponds to $1\sigma$ true-signal locations $\theta^n$ small enough to yield $r$ close to unity.
Interestingly, the signs of the frequentist and Bayesian expressions \eqref{eq:frequentistresult} and \eqref{eq:bayesianresult} are opposite, indicating (at least \emph{prima facie}) that the likelihood is overestimated in the frequentist case, underestimated in the Bayesian case. Given the conditions under which we have obtained Eqs.\ \eqref{eq:frequentistresult} and \eqref{eq:bayesianresult}, it is perhaps best to consider only their absolute value as rough indicators of the appropriateness of the high-SNR/LSA limit.

In summary, the maximum-mismatch criterion is justified from a Bayesian viewpoint by fixing the location of maximum LSA likelihood, and then exploring a surface of equiprobable true-signal locations, evaluating for each the average of $\log r$ over all experiments (i.e., realizations of noise) compatible with having the true signal there.

\subsection{Practical usage of the maximum-mismatch criterion}
\label{sec:criterionpractical}

In both the frequentist and the Bayesian pictures, Eq.\ \eqref{eq:sntest} yields the noise-averaged logarithm of the likelihood mismatch, $|\log r|$, as a function of the signal strength $A$ and of a direction in parameter space that identifies a point on the $1\sigma$ surface, given by the solutions of the LSA equation $A^2 (\bar{h}_j,\bar{h}_k) \theta_j \theta_k = 1$, and interpreted as equiprobable locations for the ML estimator given the true signal $\theta = 0$ (in the frequentist picture), or for the true signal given the mode of the likelihood at $\theta = 0$ (in the Bayesian picture).
We use Eq.\ \eqref{eq:sntest} by fixing the signal strength to what is reasonably expected in observations, perhaps close to the minimum detection SNR, although the astronomical distribution and intrinsic strengths of sources may prompt other choices (e.g., the the supermassive--black-hole binaries to be observed by LISA have typical SNRs in the hundreds); and then by evaluating $|\log r|$ as a function of direction in parameter space.

\begin{figure}
\includegraphics{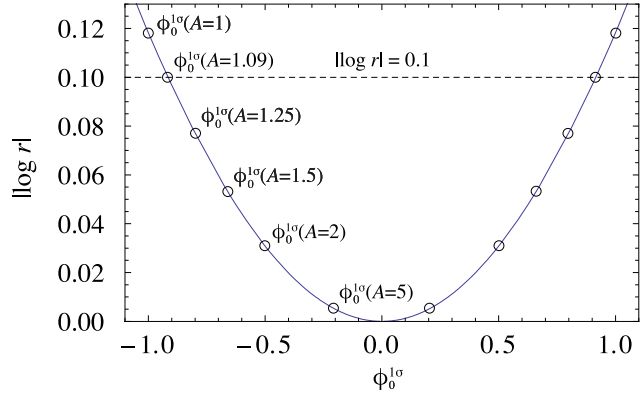}
\caption{Consistency criterion for the simple waveform model $h(\phi_0) = A \cos(2 \pi f t + \phi_0)$ in Gaussian stationary noise, with fixed (known) $A$ and $f$. The curve plots $|\log r|$ as a function of the $1\sigma$ error $\phi_0^{1\sigma} = \pm 1/A$, with specific values of $A$ called out by the circles. For the threshold $|\log r| = 0.1$, consistency is achieved for $A \gtrsim 1.09$. (To generate this graph, the integration time was set to $1000/f$, and the variance of noise adjusted so that $(\bar{h}(\phi_0),\bar{h}(\phi_0)) = 1$.)\label{fig:simplelogr}}
\end{figure}

Figure \ref{fig:simplelogr} shows an example of this procedure for a very simple and benign one-dimensional estimation problem (a sinusoid of known amplitude and frequency in Gaussian stationary noise), where the only parameter left to estimate is the initial phase $\phi_0$ ($=0$ for the true signal).
For each value of $\mathrm{SNR} \equiv A$, the expected $1\sigma$ surface consists of just the two points $\phi_0^{1\sigma} = 1/A$. Figure \ref{fig:simplelogr} shows $|\log r|$ as a function of $\phi_0^{1\sigma}$, and therefore of $A$. If we set a threshold of $|\log r| = 0.1$ (the dashed line) to claim the high-SNR/LSA limit as consistent,
we see that the consistency criterion is not satisfied for $A=1$, where $|\log r| \simeq 0.12$, but it begins to be satisfied for $A \gtrsim 1.09$. Once again, for a given SNR, $|\log r|$ at $\phi_0^{1\sigma}$ is an index of the closeness of the LSA and exact likelihoods at a typical values of the errors, and averaged among compatible noise realizations.

\begin{figure}
\includegraphics[width=5in]{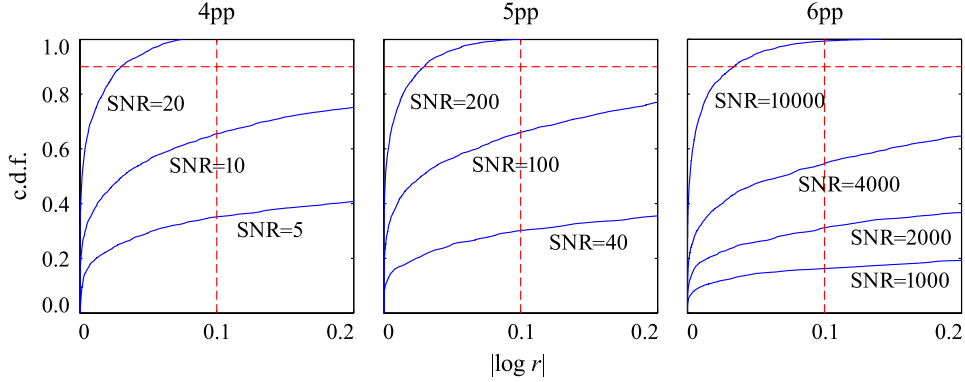}
\caption{Cumulative distribution function for $|\log r|$ on the $1 \sigma$ surface at various SNRs for our reference SPA model with $m_1 = m_2 = 10 M_\odot$. The SNR required to have 90\% of the $1\sigma$ points at $|\log r| = 0.1$ (dashed lines) increases considerably (in fact, to unrealistic values) as we move from the 4pp to 5pp and 6pp. This figure was produced without imposing any priors on the source parameters.\label{fig:logrhist}}
\end{figure}

The principle is the same for multiparameter estimation problems, where we have the additional task of sampling the entire $1\sigma$ surface in a manner consistent with the LSA distribution at $1\sigma$.
One way to do so is to obtain the eigenvalues $\lambda^{(i)}$ and eigenvectors $\theta^{(i)}_j$ of $(\bar{h}_j,\bar{h}_k)$, and then sample the parameter values $\theta = \sum_{(i)=1}^N \tilde{c}^{(i)} \theta^{(i)}_j / (A \sqrt{\lambda^{(i)}})$, with $\tilde{c}^{(i)}$ distributed uniformly on the $N$-dimensional unit sphere.\footnote{To see why this is the right thing to do, consider the integration of a function against the LSA distribution, and make a change of variables (with unit Jacobian) to eigenvalue components, and a second to rescaled components $\tilde{c}^{(i)} = A \sqrt{\lambda^{(i)}} c^{(i)}$:
\begin{displaymath}
\int (\ldots) \, e^{-A^2 \theta_i \theta_j (\bar{h}_i,\bar{h}_j) / 2} d \theta
= \int (\ldots) \, e^{-A^2 \sum_{(i)} \lambda^{(i)} [c^{(i)}]^2 / 2} d c
\propto \int (\ldots) \, e^{- \sum_{(i)} [\tilde{c}^{(i)}]^2 / 2} d \tilde{c};
\end{displaymath}
we see that the source parameters that correspond to $\tilde{c}$ lying on a sphere of fixed radius must lie on an isoprobability surface. To reassemble $\theta$ from the $\tilde{c}$, we need to \emph{divide} the eigenvectors by $A \sqrt{\lambda^{(i)}}$.} We then obtain the cumulative distribution function for the values of $|\log r|$, which we plot in Fig.\ \ref{fig:logrhist} for our reference model. If we consider the high-SNR/LSA limit to be sufficiently realized when $|\log r| < 0.1$ over 90\% of the $1\sigma$ surface, we conclude that the Fisher-matrix formalism (with no priors) is self-consistent for SNRs between 10 and 20 in the 4-parameter problem, between 100 and 200 in the 5pp, and between 4000 and 10000 in the 6pp.

\begin{figure}
\includegraphics[width=6in]{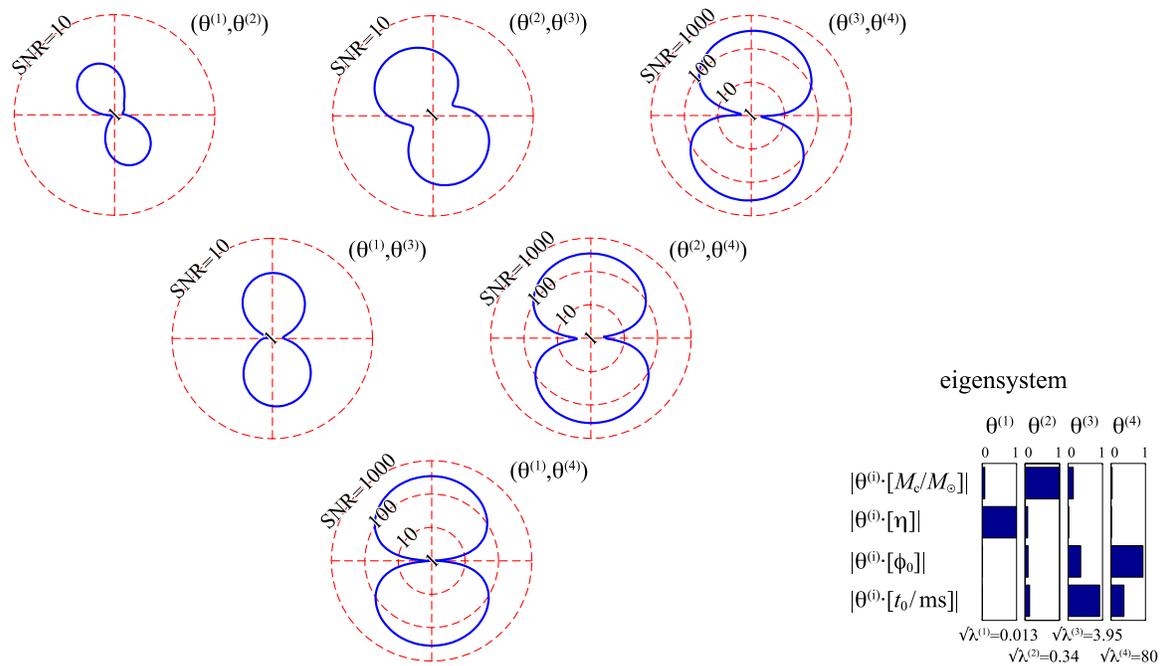}
\caption{SNR values at which $|\log r| = 0.1$ in the 4pp for our reference SPA model with $m_1 = m_2 = 10 M_\odot$. The six subplots display sections of parameter space corresponding to all distinct pairs of eigenvectors of the Fisher matrix; the polar radius of the curves shows the required SNR (plotted logarithmically from $\log \mathrm{SNR} = 0$), while the polar angle is computed between pairs of renormalized eigenvector coefficients $\tilde{c}^{(i)}$. The graph at the bottom right shows the composition of the four Fisher-matrix eigenvectors in terms of the source parameters, as well as their respective eigenvalues. This figure was produced without imposing any priors on the source parameters.\label{fig:maxsnsecs}}
\end{figure}

The eigenvector directions that push the required SNR toward higher values are usually those associated with the smallest-magnitude eigenvalues. To confirm this, and to get some clues about the beyond-LSA structure of the likelihood, we can fix 
the maximum acceptable value of $|\log r|$ (say, again to 0.1) and then solve Eq.\ \eqref{eq:sntest} for $A$ as a function of direction in parameter space. We do so for the 4pp in Fig.\ \ref{fig:maxsnsecs}, where we show all two-dimensional parameter subspaces along pairs of eigenvectors (strictly speaking, were are not sampling a single $1\sigma$ surface, but considering the set of such surfaces for all SNRs, and determining on which of them $|\log r| = 0.1$, as a function of parameter angle).

In the Bayesian framework, it is also possible to combine the maximum-mismatch criterion with the normal prior test of Sec.\ \ref{sec:priorexample}, by investigating the values of $|\log r|$ on the $1\sigma$ surface given by the solutions of the LSA-\emph{cum}-\emph{prior} equation $[(h_j,h_k) + P_{jk}]\theta_j \theta_k = 1$. This test can help decide whether the LSA is warranted once priors are factored in: it can be shown that Eq.\ \eqref{eq:bayesianresult} continues to hold with NTC priors, although its interpretation is not as clean, because their mode moves around the $1\sigma$ surface as we explore it. The maximum-mismatch criterion may indicate that, at the signal strengths of interest, the LSA becomes consistent only with the priors; in that case, the reliable predictor of source parameter accuracy would not be $[(h_j,h_k) + P_{jk}]^{-1}$ (given the crudeness of NTC priors), but rather the result of an LSA Monte Carlo procedure such as that described in Sec.\ \ref{sec:exactprior}.

It is very hard to make a general statement about the errors in the expected accuracies when the LSA Fisher-matrix result is not self-consistent. Such errors are strongly SNR-dependent, and it is usually necessary to include parameter priors into consideration. As anecdotal evidence, I offer that
for our reference model at $\mathrm{SNR} = 10$, a full-blown Monte Carlo sampling of the posterior distribution, involving an explicit time-domain realization of noise and adopting the priors of Sec.\ \ref{sec:exactprior}, reports posterior variances that differ from the last three rows of Tab.\ \ref{tab:priors1010} by few tens percent for the 4pp, and by factors of a few (for $M_c$ and $\eta$ only, since $\Delta \beta$ and $\Delta \sigma$ are dominated by the priors) for the 5pp and 6pp.

In conclusion, I submit that graphs like those of Fig.\ \ref{fig:logrhist} can be useful to assess the consistency of the ``straight'' Fisher-matrix formalism, are easy to produce with little additional machinery, and should be included in all articles that use the formalism to predict the future parameter-estimation performance of GW observations. If a single number must be quoted, it could be the SNR at which 90\% of the $1\sigma$ surface yields $|\log r| < 0.1$.

\section{Beyond the linearized-signal approximation}
\label{sec:higherorder}

In this section we develop mathematical tools to derive higher-than-LSA expressions for the frequentist mean and variance of the ML estimator over an ensemble of noise realizations (Sec.\ \ref{sec:higherfrequentist}), and for the Bayesian mean and variance of the posterior distribution (without priors) in a single experiment (Sec.\ \ref{sec:higherbayesian}). These expressions provide corrections to the Fisher-matrix result, and can therefore be used to check its accuracy, as suggested by Cutler and Flanagan \cite{cf1994}, who derive a general expression for the $1/A^4$ correction to the frequentist variance. A formal treatment of the $1/A$ expansion for the frequentist moments can be found in Barndorff-Nielsen and Cox \cite{bnc1994} and in Zanolin, Naftali, and Makris \cite{znm2007}, who use the expansion to determine conditions for the ML estimate to become unbiased and attain the Cram\'er--Rao bound \cite{nm2001}.

However, computing higher-order corrections involves a considerable amount of tensorial algebra that calls for the use of specialized software, such as \textsc{MathTensor} \cite{mathtensor}; they also involve higher-than-first derivatives of the waveforms and products of several inverse Fisher matrices, which may raise concerns about the numerical accuracy of the computations.
Throughout this section, we distinguish between covariant and contravariant indices (as in $n_i = (\bar{h}_i,n)$ and $\theta^i$, respectively); in fact, we find it convenient to use the inverse normalized Fisher matrix $(\bar{h}^i,\bar{h}^j)^{-1}$ to raise indices, therefore hiding its repeated appearance in tensor expressions.

\subsection{In the frequentist framework}
\label{sec:higherfrequentist}

Using the $1/A$ expansion of Eqs.\ \eqref{eq:mlexpanded} and \eqref{eq:mlestimatorexpand}, the perturbative ML equations can be written in general as
\begin{equation}
\begin{aligned}
H_{ij} \theta^j_{(1)} & = N_i, && \\
H_{ij} \theta^j_{(2)} & = N_{ij} \theta^j_{(1)} 
& -\,& H_{ijk} \theta^{jk}_{(2)}, \\
H_{ij} \theta^j_{(3)} & = N_{ij} \theta^j_{(2)} + N_{ijk} \theta^{jk}_{(2)}
& -\,& H_{ijk} \theta^{jk}_{(3)} - H_{ijkl} \theta^{jkl}_{(3)}, \\
H_{ij} \theta^j_{(4)} & = N_{ij} \theta^j_{(3)} + N_{ijk} \theta^{jk}_{(3)} + N_{ijkl} \theta^{jkl}_{(3)} \!\!\!\!\!
& -\,& H_{ijk} \theta^{jk}_{(4)} - H_{ijkl} \theta^{jkl}_{(4)} - H_{ijklm} \theta^{jklm}_{(3)}, \\
& \cdots &&
\end{aligned}
\label{eq:mleqsexpand}
\end{equation}
with $N_i = (n,\bar{h}_i) / 0!$, $N_{ij} = (\bar{h}_{ij},n) / 1!$, $N_{ijk} = (\bar{h}_{ijk},n) / 2!$ (and so on), and
\begin{equation}
\begin{aligned}
H_{ij} & = \frac{\bar{h}_{i,j}}{0! \, 1!}, \\
H_{ijk} & = \frac{\bar{h}_{i,jk}}{0! \, 2!} + \frac{\bar{h}_{ij,k}}{1! \, 1!}, \\
H_{ijkl} & = \frac{\bar{h}_{i,jkl}}{0! \, 3!} + \frac{\bar{h}_{ij,kl}}{1! \, 2!} + \frac{\bar{h}_{ijk,l}}{2! \, 1!}, \\
& \cdots
\end{aligned}
\label{eq:defbigh}
\end{equation}
where $\bar{h}_{i,j} = (\bar{h}_i,\bar{h}_j)$, $\bar{h}_{i,jk} = (\bar{h}_i,\bar{h}_{jk})$ (and so on), and where the two factorials at each denominator are those (respectively) of the number of indices before the comma minus one, and of the number of indices after the comma. Also, the $\theta^j_{(n)}$ of Eq.\ \eqref{eq:mleqsexpand} are the unknown $1/A^n$ contributions to $\hat{\theta}^\mathrm{ML}$ (as in Eq.\ \eqref{eq:mlestimatorexpand}, dropping hats for simplicity), while the multi-index parameter objects such as $\theta^{jk}_{(2)}$ are given by
\begin{equation}
\begin{array}{lll}
\theta^{jk}_{(2)} = \theta^j_{(1)} \theta^k_{(1)}, \quad &
\theta^{jk\phantom{l}}_{(3)} = \theta^j_{(1)} \theta^k_{(2)} + \theta^j_{(2)} \theta^k_{(1)}, \quad &
\theta^{jk\phantom{lm}}_{(4)} = \theta^j_{(1)} \theta^k_{(3)} + \theta^j_{(2)} \theta^k_{(2)} + \theta^j_{(3)} \theta^k_{(1)}, \\[2mm]
&
\theta^{jkl}_{(3)} = \theta^j_{(1)} \theta^k_{(1)} \theta^l_{(1)}, &
\theta^{jkl\phantom{m}}_{(4)} = \theta^j_{(1)} \theta^k_{(1)} \theta^l_{(2)} +
                     \theta^j_{(1)} \theta^k_{(2)} \theta^l_{(1)} +
                     \theta^j_{(2)} \theta^k_{(1)} \theta^l_{(1)}, \\[2mm]
& &
\theta^{jklm}_{(4)} = \theta^j_{(1)} \theta^k_{(1)} \theta^l_{(1)} \theta^m_{(1)},
\end{array}
\end{equation}
and so on. In general, the object $\theta^{j_1{\cdots}j_m}_{(n)}$ will consist of as many addends as there are partitions of $n$ into $m$ integers, including all permutations of each partition. For instance, the $n=5$, $m=3$ object $\theta_{(5)}^{jkl}$ would have terms for each of the partitions $1+1+3$, $1+3+1$, $3+1+1$, $1+2+2$, $2+1+2$, $2+2+1$.

The solution of each equation in Eq.\ \eqref{eq:mleqsexpand} is trivial given the solutions of all equations of lower order. Since the inverse matrix $(H^{-1})^{ij} \equiv (\bar{h}_i,\bar{h}_j)^{-1} = A^2 F_{ij}^{-1}$ appears multiple times in the solutions (because $H_{ij}$ multiplies the unknown $\theta^j_{(n)}$ in each equation), it is convenient to adopt a compact notation that hides the $(H^{-1})^{ij}$ by raising every index into which they are contracted. We then find
\begin{equation}
\begin{aligned}
\theta_{(1)}^i & = N^i, \\
\theta_{(2)}^i & =
{N^i}_j \theta_{(1)}^j - {H^i}_{jk} \theta_{(2)}^{jk} =
{N^i}_j N^j - {H^i}_{jk} N^j N^k, \\
\theta_{(3)}^i & = {N^i}_j \theta_{(2)}^j + {N^i}_{jk} \theta_{(1)}^j \theta_{(1)}^k
- {H^i}_{jk} (\theta_{(1)}^j \theta_{(2)}^k + \theta_{(2)}^j \theta_{(1)}^k) - {H^i}_{jkl} \theta_{(1)}^j \theta_{(1)}^k \theta_{(1)}^l = \cdots \\
& \ldots
\end{aligned}
\end{equation}

The frequentist mean and covariance of the $\theta^i$ can be built from these solutions, remembering the Wick-product rule \cite{zinnjustin} for the ensemble average of products of Gaussian variables:
\begin{equation}
\begin{aligned}
\langle(a,n)\rangle_n & = 0, \\
\langle(a,n)(b,n)\rangle_n & = (a,b), \\
\langle(a,n)(b,n)(c,n)\rangle_n & = 0, \\
\langle(a,n)(b,n)(c,n)(d,n)\rangle_n & = (a,b)(c,d) + (a,c)(b,d) + (a,d)(b,c) \\
& \ldots
\end{aligned}
\label{eq:wick}
\end{equation}
(for any signals $a$, $b$, $c$, and $d$), where all the products with an odd number of factors vanish, while the products with an even number of factors are given by the sum of terms corresponding to all distinct pairings of signals into inner products.
Thus we find that all the $\langle \theta^i_{\mathrm{odd}\,(k)} \rangle_n$ vanish, while the first non-zero correction to $\langle \theta^i \rangle_n$ is
\begin{equation}
\langle \theta_{(2)}^i \rangle_n = \langle {N^i}_j N^j \rangle_n - {H^i}_{jk} \langle N^j N^k \rangle_n = \bar{h}^{ij}_{\phantom{ij},j} - {H^i}_{jk} \bar{h}^{j,k}.
\end{equation}
As for the covariance,
\begin{multline}
\langle \theta^i \theta^j \rangle_n
- \langle \theta^i \rangle_n \langle \theta^j \rangle_n
= \biggl[
\frac{1}{A^2} \bigl\langle \theta^i_{(1)} \theta^j_{(1)} \bigr\rangle_n +
\frac{1}{A^3} \bigl\langle \cancel{\theta^i_{(1)} \theta^j_{(2)}} + \cancel{\theta^i_{(2)} \theta^j_{(1)}} \bigr\rangle_n +
\frac{1}{A^4} \bigl\langle \theta^i_{(1)} \theta^j_{(3)} + \theta^i_{(2)} \theta^j_{(2)} + \theta^i_{(3)} \theta^j_{(1)} \bigr\rangle_n + \cdots \biggr] \\ 
- \biggl[
\frac{1}{A^2} \bigl\langle \cancel{\theta^i_{(1)}} \rangle_n \langle \cancel{\theta^j_{(1)}} \bigr\rangle_n + \frac{1}{A^3} \Bigl(
 \bigl\langle \cancel{\theta^i_{(1)}} \bigr\rangle_n \langle \theta^j_{(2)} \bigr\rangle_n +
 \bigl\langle \theta^i_{(2)} \bigr\rangle_n \langle \cancel{\theta^j_{(1)}} \bigr\rangle_n
\Bigr) \\
+ \frac{1}{A^4} \Bigl(
 \bigl\langle \cancel{\theta^i_{(1)}} \bigr\rangle_n \langle \cancel{\theta^j_{(3)}} \bigr\rangle_n +
 \bigl\langle \theta^i_{(2)} \bigr\rangle_n \langle \theta^j_{(2)} \bigr\rangle_n +
 \bigl\langle \cancel{\theta^i_{(3)}} \bigr\rangle_n \langle \cancel{\theta^j_{(1)}} \bigr\rangle_n
\Bigr) + \cdots
\biggr],
\label{eq:fourthcovariance}
\end{multline}
where all the stricken-through terms vanish because they are proportional to ensemble products of an odd number of $n$ terms. The surviving contributions are given by
\begin{equation}
\begin{aligned}
\bigl\langle \theta^i_{(1)} \theta^j_{(1)} \bigr\rangle_n =\, & 
\bigl\langle N^i N^j \bigr\rangle_n = \bar{h}^{i,j}, \\
\bigl\langle \theta^i_{(1)} \theta^j_{(3)} \bigr\rangle_n =\, &
\bigl\langle N^i {N^j}_k {N^k}_l N^l \bigr\rangle_n -
{H^k}_{lm} \bigl\langle N^i {N^j}_k N^l N^m \bigr\rangle_n
+ \bigl\langle N^i {N^j}_{kl} N^k N^l \bigr\rangle_n - {H^j}_{klm} \bigl\langle N^i N^k N^l N^m \bigr\rangle_n \\
& - {H^j}_{kl} \Bigl(
\bigl\langle N^i N^k {N^l}_m N^m \bigr\rangle_n
- {H^l}_{mq} \bigl\langle N^i N^k N^m N^q \bigr\rangle_n + \bigl\langle N^i {N^k}_m N^m N^l \bigr\rangle_n
- {H^k}_{mq} \bigl\langle N^i N^m N^q N^l \bigr\rangle_n
\Bigr), \\
\bigl\langle \theta^i_{(2)} \theta^j_{(2)} \bigr\rangle_n =\, &
\bigl\langle {N^i}_k N^k {N^j}_m N^m \bigr\rangle_n
- {H^i}_{kl} \bigl\langle N^k N^l {N^j}_m N^m \bigr\rangle_n
- {H^j}_{mq} \bigl\langle N^m N^q {N^i}_k N^k \bigr\rangle_n
+ {H^i}_{kl} {H^j}_{mq} \bigl\langle N^k N^l N^m N^q \bigr\rangle_n,
\end{aligned}
\label{eq:lotsofn}
\end{equation}
and of course $\langle \theta^i_{(3)} \theta^j_{(1)} \rangle_n = \langle \theta^j_{(1)} \theta^i_{(3)} \rangle_n$.
The first of these equations reproduces the standard Fisher-matrix result.
The four-$N$ products in Eq.\ \eqref{eq:lotsofn} follow from Eq.\ \eqref{eq:wick}. For instance, the last two products are given by
\begin{equation}
\begin{aligned}
\bigl\langle N^k N^l N^m N^q \bigr\rangle_n &=
\bigl\langle N^k N^l \bigr\rangle_n \bigl\langle N^m N^q \bigr\rangle_n +
\bigl\langle N^k N^m \bigr\rangle_n \bigl\langle N^l N^q \bigr\rangle_n +
\bigl\langle N^k N^q \bigr\rangle_n \bigl\langle N^l N^m \bigr\rangle_n \\
& = \bar{h}^{k,l}\bar{h}^{m,q} + \bar{h}^{k,m}\bar{h}^{l,q} + \bar{h}^{k,q}\bar{h}^{l,m}, \\
\bigl\langle N^m N^q {N^i}_k N^k \bigr\rangle_n &=
\bar{h}^{m,q} \, \bar{h}^{ik,}_{\phantom{ik,}k} +
\bar{h}^{m,i}_{\phantom{m,i}k} \, \bar{h}^{q,k} +
\bar{h}^{m,k} \, \bar{h}^{q,i}_{\phantom{q,i}k} .
\end{aligned}
\label{eq:nproducts}
\end{equation}
These expressions can be substituted into those of Eq.\ \eqref{eq:lotsofn}, and those into Eq.\ \eqref{eq:fourthcovariance}, yielding the frequentist variance to order $1/A^4$. Unfortunately, this requires computing second- and third-order waveform derivatives (the latter for $H_{jklm}$).

\subsection{In the Bayesian framework}
\label{sec:higherbayesian}

To generalize Eq.\ \eqref{eq:bayesianseries}, we write
\begin{equation}
\mathcal{I}^{(n)} = \int \bar{\theta}^i \frac{\partial^n p(0)}{\partial \epsilon^n} d\theta,
\quad
\mathcal{N}^{(n)} = \int \frac{\partial^n p(0)}{\partial \epsilon^n} d\theta,
\end{equation}
and find the recurrence relation
\begin{equation}
\langle \bar{\theta}^i \rangle_p^{(n)} =
\biggl(\mathcal{I}^{(n)} - \sum_{j=1}^n \biggl[ \biggl(\begin{array}{c} n \\ j \end{array} \biggr) \langle \bar{\theta}_i \rangle_p^{(n-j)} \times \mathcal{N}^{(j)}
\biggr] \biggr) \Big/ \mathcal{N}^{(0)},
\label{eq:recurrence}
\end{equation}
which we may prove by expanding the identity $\mathcal{I}^{(0)} + \epsilon \, \mathcal{I}^{(1)} + \frac{\epsilon^2}{2} \mathcal{I}^{(2)} + \cdots =
[\langle \bar{\theta}^i \rangle_p^{(0)} + \epsilon \langle \bar{\theta}^i \rangle_p^{(1)} + \cdots] \times [\mathcal{N}^{(0)} + \epsilon \, \mathcal{N}^{(1)} + \cdots]$ on both sides as a series of $\epsilon$, leading to
\begin{equation}
\mathcal{I}^{(n)} = \sum_{j=0}^n \biggl(\begin{array}{c} n \\ j \end{array} \biggr) \mathcal{N}^{(j)} \langle \bar{\theta}_i \rangle_p^{(n-j)},
\end{equation}
whence Eq.\ \eqref{eq:recurrence}. To obtain all needed derivatives with respect to $\epsilon$, we rewrite Eq.\ \eqref{eq:anotherp} as
\begin{equation}
\begin{aligned}
p(s|\theta) \propto \exp \Bigl\{ & -(n,n)/2 +
\Bigl[ N_i \bar{\theta}^i + \epsilon N_{ij} \bar{\theta}^i \bar{\theta}^j +
\epsilon^2 N_{ijk} \bar{\theta}^i \bar{\theta}^j \bar{\theta}^k + \cdots \Bigr] \\
& - \Bigl[
H_{jk} \bar{\theta}^j \bar{\theta}^k +
\epsilon H'_{jkl} \bar{\theta}^j \bar{\theta}^k \bar{\theta}^l +
\epsilon^2 H'_{jklm} \bar{\theta}^j \bar{\theta}^k \bar{\theta}^l \bar{\theta}^m + \cdots
\Bigr] / 2 \Bigr\},
\end{aligned}
\end{equation}
where $N_i = (n,\bar{h}_i) / 1! = n_i$, $N_{ij} = (\bar{h}_{ij},n) / 2! = n_{ij}/2$, $N_{ijk} = (\bar{h}_{ijk},n) / 3!$ (and so on), and also the $H'_{j_1{\cdots}j_n}$ have slightly different denominators than the $H_{j_1{\cdots}j_n}$ of Eq.\ \eqref{eq:defbigh}:
\begin{equation}
\begin{aligned}
H'_{ijk} & = \frac{\bar{h}_{i,jk}}{1! \, 2!} + \frac{\bar{h}_{ij,k}}{2! \, 1!}, \\
H'_{ijkl} & = \frac{\bar{h}_{i,jkl}}{1! \, 3!} + \frac{\bar{h}_{ij,kl}}{2! \, 2!} + \frac{\bar{h}_{ijk,l}}{3! \, 1!}; \\
& \cdots
\end{aligned}
\end{equation}
namely, the denominator is $m! \, l!$ for the product $h_{j_1{\cdots}j_m,j_1{\cdots}j_l} \equiv (\bar{h}_{j_1{\cdots}j_m},\bar{h}_{j_1{\cdots}j_l})$. Expanding as a series of $\epsilon$ yields
\begin{equation}
\begin{aligned}
p(s|\theta) \propto e^{-H_{jk} \bar{\theta}^j \bar{\theta}^k / 2 + n_j \bar{\theta}^j} \times
\biggl\{
1 & + \epsilon \Bigl( N_{jk} \bar{\theta}^j \bar{\theta}^k
- \frac{1}{2} H'_{jkl} \bar{\theta}^j \bar{\theta}^k \bar{\theta}^l \Bigr) \\
& + \frac{\epsilon^2}{2} \biggl(
\Bigl( N_{jk} \bar{\theta}^j \bar{\theta}^k
- \frac{1}{2} H'_{jkl} \bar{\theta}^j \bar{\theta}^k \bar{\theta}^l \Bigr)^2
+ 2 \Bigl(
N_{jkl} \bar{\theta}^j \bar{\theta}^k \bar{\theta}^l
- \frac{1}{2} H'_{jklm} \bar{\theta}^j \bar{\theta}^k \bar{\theta}^l \bar{\theta}^m \Bigr) \biggr) + \cdots
\biggr\},
\end{aligned}
\end{equation}
so that the $\mathcal{I}^{(n)}$ and $\mathcal{N}^{(n)}$ are given by expressions akin to
\begin{equation}
\mathcal{I}^{(1)} / \mathcal{N}^{(0)} = 
\int \bar{\theta}^i \Bigl( N_{jk} \bar{\theta}^j \bar{\theta}^k
- \frac{1}{2} H'_{jkl} \bar{\theta}^j \bar{\theta}^k \bar{\theta}^l \Bigr)
e^{-H_{jk} \bar{\theta}^j \bar{\theta}^k / 2 + n_j \bar{\theta}^j} d \theta
\bigg/\!\! \int e^{-H_{jk} \bar{\theta}^j \bar{\theta}^k / 2 + n_j \bar{\theta}^j} d \theta.
\end{equation}
Now, the integrals of the general form
\begin{equation}
\langle \bar{\theta}^{i_1} \cdots \bar{\theta}^{i_m} \rangle_p^{(0)} = 
\int \bar{\theta}^{i_1} \cdots \bar{\theta}^{i_m}
e^{-H_{jk} \bar{\theta}^j \bar{\theta}^k / 2 + n_j \bar{\theta}^j} d \theta
\bigg/\!\! \int e^{-H_{jk} \bar{\theta}^j \bar{\theta}^k / 2 + n_j \bar{\theta}^j} d \theta,
\label{eq:intgenform}
\end{equation}
can be computed with the Wick identity\footnote{Another way to organize this computation is to offset the integration variable $\bar{\theta}^j$ to $\bar{\theta}^j - (H^{jk})^{-1} n_k = \bar{\theta}^j - n^j$ in Eq.\ \eqref{eq:intgenform}, obtaining
\begin{displaymath}
\langle \bar{\theta}^{i_1} \cdots \bar{\theta}^{i_m} \rangle_p^{(0)} = 
\int (\bar{\theta}^{i_1} + n^{i_1}) \cdots (\bar{\theta}^{i_m} + n^{i_m})
e^{-H_{jk} \bar{\theta}^j \bar{\theta}^k / 2} d \theta
\bigg/\!\! \int e^{-H_{jk} \bar{\theta}^j \bar{\theta}^k / 2} d \theta;
\end{displaymath}
we can then expand the product in the integrand, bring the $n^{i_k}$ outside the integral, and apply Wick's theorem [Eq.\ \eqref{eq:wick}] to obtain each addend of the form
\begin{displaymath}
n^{i_1} \cdots n^{i_{m-l}} \int \bar{\theta}^{i_1} \cdots \bar{\theta}^{i_l}
e^{-H_{jk} \bar{\theta}^j \bar{\theta}^k / 2} d \theta
\bigg/\!\! \int e^{-H_{jk} \bar{\theta}^j \bar{\theta}^k / 2} d \theta;
\end{displaymath}
all integrals with odd $l$ are zero, while the integrals with even $l$ are given by the sum of all possible pairings of indices into products of $(H^{\cdots})^{-1}$.} \cite{zinnjustin}
\begin{equation}
\bigl\langle F(\bar{\theta}) \bigr\rangle_p^{(0)} = F\biggl(\frac{\partial}{\partial n}\biggr) \exp \bigl\{n_i (H^{ij})^{-1} n_j / 2\bigr\};
\end{equation}
in particular (again using $(H^{ij})^{-1}$ to raise indices),
\begin{equation}
\begin{aligned}
\langle \bar{\theta}^i \rangle^{(0)}_p & = n^i, \\
\langle \bar{\theta}^i \bar{\theta}^j \rangle^{(0)}_p & =
(H^{ij})^{-1} + n^i n^j, \\
\langle \bar{\theta}^i \bar{\theta}^j \bar{\theta}^k \rangle^{(0)}_p & =
(H^{ij})^{-1} n^k + (H^{ik})^{-1} n^j + n^i (H^{jk})^{-1} + n^i n^j n^k. \\
& \ldots
\end{aligned}
\end{equation}

Unfortunately, the $1/A^4$ (i.e., $\epsilon^2$) corrections to the variance turn out to be rather unwieldy, and belong in a symbolic-manipulation software package rather than on these pages.
We content ourselves with the $1/A^2$ correction to the posterior mean (remember that the normalized parameters $\bar{\theta}$ carry an $A$),
\begin{equation}
\langle \bar{\theta}^i \rangle_p = n^i + \epsilon
\Bigl[
{n^i}_k n^k - \bigl(
\tfrac{1}{2} \bar{h}^{i,}_{\phantom{i,}kl} + \bar{h}^{i}_{\phantom{i}k,l}
\bigr) \bigl(
n^k n^l + \bar{h}^{kl}
\bigr)
\Bigr] + O(\epsilon^2),
\end{equation}
and the $1/A^3$ correction to the variance,
\begin{multline}
\langle \bar{\theta}^i \bar{\theta}^j \rangle_p
- \langle \bar{\theta}^i \rangle_p
\langle \bar{\theta}^j \rangle_p
= \bar{h}^{ij} + \epsilon
\Bigl[
n^{ij} + \tfrac{1}{2} n^i n^j n_{kl} \bigl( n^k n^l - \bar{h}^{kl} \bigr)
- n^k \bigl( \bar{h}^{i,j}_{\phantom{i,j}k} + \bar{h}^{j,i}_{\phantom{j,i}k} + \bar{h}_{k,}^{\phantom{k,}ij} \bigr)
\\ - n^i n^j n^k \bigl( \tfrac{1}{2} \bar{h}_{k,l}^{\phantom{k,l}l}
+ \bar{h}^{l,}_{\phantom{l,}kl} + \tfrac{1}{2} \bar{h}_{k,lm} n^l n^m
\bigr)
\Bigr] + O(\epsilon^2).
\end{multline}

Thus we see that the $1/A^3$ contribution to the variance does not vanish in any single experiment (unless $n^i = 0$). It does vanish, however, under frequentist average, since it involves products of odd numbers of noises.

\section{Conclusion}

In this article I tried to provide, as it were, a user's manual for the Fisher information matrix. It seems clear that the Fisher-matrix formalism will continue to be featured prominently in research dealing with the parameter-estimation prospects of future GW observations, because of its compactness and accessibility, and because of the difficulty of computing higher-order corrections and running full-blown simulations. Yet the three questions posed in the introduction loom over the credibility of Fisher-matrix results, which is all the more worrisome when these results are used to justify choices in science policy or experiment design.

The recipes provided in this paper to answer the initial questions can help assert (or falsify) the accuracy of the formalism for specific signal models. In particular:
\begin{enumerate}
\item As discussed in Sec.\ \ref{sec:disappearing}, ill-conditioned or singular Fisher matrices point to the need for increased numerical accuracy, and occasionally to a case for discarding a parameter or combination of parameters, but more often to suspicions about the appropriateness of the high-SNR/LSA limit. Section \ref{sec:singularfisher} describes how to use the singular value decomposition of the Fisher matrix to discard truly degenerate linear combinations of parameters; Sec.\ \ref{sec:approxfisher} describes how to roughly assess the sensitivity of the Fisher-matrix inverse to numerical error by means of the Fisher-matrix condition number, and more carefully by a simple Monte Carlo test.
\item The necessity of including prior distributions for the source parameters, perhaps in as simple a form as uniform distributions over the physically allowed ranges, can be roughly assessed by verifying whether Fisher-matrix results change with the addition of simple Gaussian priors, as shown in Sec.\ \ref{sec:priorexample}; more accurate estimates of the effect of priors can be obtained by integrating the variance of an exact-prior--LSA-likelihood posterior with the simple Monte Carlo algorithm of Sec.\ \ref{sec:exactprior}.
\item The detected-signal strength (i.e., the SNR) necessary for Fisher-matrix results to be internally consistent can be evaluated with the likelihood-mismatch criterion that follows from Eqs.\ \eqref{eq:frequentistresult} and \eqref{eq:bayesianresult} of Sec.\ \ref{sec:howhigh}, or (at the price of some algebra) by computing the higher-order corrections presented in Sec.\ \ref{sec:higherorder}. 
\end{enumerate}
If the Fisher-matrix formalism remains inconsistent at the SNRs of interest, even with the help of priors, there is little recourse but to embark in explicit Monte Carlo simulations of frequentist \cite{bsd1996} or Bayesian \cite{mcmc} parameter estimation. Such simulations can consistently include sophisticated priors, and explore the secondary maxima of the posterior (or likelihood, in the frequentist case). They are the gold standard of this trade, but as such they are expensive in human effort and CPU resources. The recipes given in this paper can help establish when they are truly needed.

\acknowledgments

I would like to thank Yanbei Chen, Curt Cutler, Yi Pan, and Michele Zanolin for useful discussions; for reviewing this manuscript, I am grateful to John Armstrong, Emanuele Berti, Steve Drasco, Frank Estabrook, Sam Finn, \'Eanna Flanagan, and especially Alessandra Buonanno. My work was supported by the LISA Mission Science Office and by the Human Resources Development Fund at the Jet Propulsion Laboratory, California Institute of Technology, where it was performed under contract with the National Aeronautics and Space Administration.

\appendix

\section{Lemma for the conditional average of jointly normal random variables}
\label{sec:conditional}

Assume the vector $x_j$ and the scalar $y$ are jointly normal random variables with mean zero and covariance matrix
\begin{equation}
\mathcal{C} = \left(
\begin{array}{cc}
F_{ij} & H_{i} \\
H_{j}  & G
\end{array}
\right).
\label{eq:jointnormal}
\end{equation}
From the standard Frobenius--Schur formula for the inverse of a block matrix \cite{bodewig},
\begin{equation}
\label{eq:schur}
\left(
\begin{array}{cc}
A & B \\
C & D
\end{array}
\right)^{-1} = 
\left(
\begin{array}{cc}
A^{-1} + A^{-1} B S^{-1}_A C A^{-1} & -A^{-1} B S^{-1}_A \\
-S^{-1}_A C A^{-1} & S^{-1}_A
\end{array}
\right)
\end{equation}
(with $S_A = D - C A^{-1} B$ the \emph{Schur complement} of $A$),
we find
\begin{equation}
\mathcal{C}^{-1} =
\left(
\begin{array}{cc}
F^{-1}_{ij} + S^{-1}_A (F^{-1}_{ik} H_k) (F^{-1}_{jl} H_l) &
-S^{-1}_A (F^{-1}_{jk} H_k) \\
-S^{-1}_A (F^{-1}_{ik} H_k) & S^{-1}_A
\end{array}
\right),
\end{equation}
since in our case $F^{-1}_{ij}$ is symmetric and $S_A$ is the scalar $G - (F^{-1}_{ij} H_i H_j)$. Now, the joint distribution of $x_j$ and $y$ is given by
\begin{equation}
p(x,y) \propto \exp - \left\{ \left( x_i \; y \right) \cdot
\mathcal{C}^{-1}
\cdot \left( \begin{array}{c} x_i \\ y \end{array} \right)
\right\} / \, 2,
\end{equation}
while the conditional distribution of $y$ given $x_j$ is $p(y|x) = p(x,y)/p(x) = p(x,y) / \left[\int p(x,y) \, dy\right]$. Since however $p(x)$ can be a function only of $x$, by the properties of Gaussian integrals it must be that $p(x) \propto \exp \, (\cdots)_{ij} x_i x_j$. It follows that $p(y|x)$ must be of the form
\begin{equation}
p(y|x) \propto \exp - \left\{
S_A^{-1} y^2 - 2 S_A^{-1} (x_i F_{ij}^{-1} H_j) y + (\cdots)_{ij} x_i x_j
\right\} / \, 2,
\end{equation}
from which, by inspection, we conclude that
\begin{equation}
\langle y \rangle_{x_i} = \int y \, p(y|x)\, dy = x_i F_{ij}^{-1} H_j
\end{equation}
and that
\begin{equation}
\mathrm{var}_{x_i} y = \int (y - \langle y \rangle_{x_i})^2 \, p(y|x) \, dx =
S_A = G - F_{ij}^{-1} H_i H_j.
\end{equation}

\end{document}